\newif\ifsubmode
\submodefalse

\newif\ifprintfig
\printfigtrue

\newif\ifemulate
\emulatetrue
 \ifsubmode
\documentclass[12pt,preprint]{aastex}
 \received{}
 \accepted{}
 \journalid{}
 \articleid{}
 \else
    \documentclass{emulateapj}
    \submitted{{\it Accepted for publication in ApJ}}
 \fi
 \usepackage{multirow,color,natbib,wrapfig,ulem}
\newcommand{\jcap}{{\it JCAP}}
\newcommand{\pasa}{{\it PASA}}
\def\spose#1{\hbox to 0pt{#1\hss}}
\def\simlt{\mathrel{\spose{\lower 3pt\hbox{$\mathchar"218$}}
     \raise 2.0pt\hbox{$\mathchar"13C$}}}
\def\simgt{\mathrel{\spose{\lower 3pt\hbox{$\mathchar"218$}}
     \raise 2.0pt\hbox{$\mathchar"13E$}}}

\shorttitle{Satellite dynamics}
\shortauthors{Brooks \& Zolotov}

\begin{document} 

\title{Why Baryons Matter: The Kinematics of Dwarf Spheroidal Satellites}

\author{Alyson\ M.\ Brooks\altaffilmark{1,2}, Adi\ Zolotov\altaffilmark{3}}

\altaffiltext{1}{Department of Astronomy, University of Wisconsin-Madison, 475 N. Charter St., Madison, WI 53706} 
\altaffiltext{2}{Department of Physics \& Astronomy, Rutgers, The State University of New Jersey, 136 Frelinghuysen Rd,
Piscataway, NJ 08854; abrooks@physics.rutgers.edu}
\altaffiltext{3}{Racah Institute of Physics, The Hebrew University,
  Jerusalem, Israel 91904; adizolotov@gmail.com}

\date{\today}

\begin{abstract}
We use high resolution 
cosmological simulations of Milky Way-mass galaxies that include 
both baryons and dark matter to show that baryonic physics (energetic feedback 
from supernovae and subsequent tidal stripping) significantly reduces the dark 
matter mass in the central regions of luminous satellite galaxies.  The reduced 
central masses of the simulated satellites reproduce the observed internal dynamics 
of Milky Way and M31 satellites as a function of luminosity.  We use these 
realistic satellites to update predictions for the observed velocity and luminosity 
functions of satellites around Milky Way-mass galaxies when baryonic effects are 
accounted for.  We also predict that field dwarf galaxies in the same luminosity 
range as the Milky Way classical satellites should not exhibit velocities as low 
as the satellites, since the field dwarfs do not experience tidal stripping.  
Additionally, the early formation times of the satellites compared to field galaxies 
at the same luminosity may be apparent in the star formation histories of the 
two populations.  Including baryonic 
physics in Cold Dark Matter models naturally explains the observed low dark matter 
densities in the Milky Way's dwarf spheroidal population.  Our simulations therefore  
resolve the tension between kinematics predicted in Cold Dark Matter theory and 
observations of satellites, without invoking alternative forms of dark matter.
\end{abstract}

\section{Introduction}
There are fewer small satellite galaxies orbiting our Milky Way (MW) galaxy
than predicted by the favored Cold Dark Matter (CDM) cosmological model  
\citep{Moore1999, Klypin1999, Madau2008, Wadepuhl2011, Brooks2013}.  Theories 
often reconcile the discrepancy between the number of observed satellites and 
CDM predictions by invoking the suppression of star formation in low mass 
galaxies, for example by UV heating at reionization 
\citep[e.g.,][]{Okamoto2008}.  If only the most massive satellites 
form stars, this can bring the predicted number of luminous satellites down 
from thousands to tens, in line with observations.  Even then a serious problem 
remains, as the most massive satellites predicted by CDM models are still much 
too dense compared to what we observe \citep{Boylan-kolchin2011, 
Boylan-kolchin2012, Wolf2012, Hayashi2012, Tollerud2012, Martinez2013}. The tension between 
the observations and the predictions of the CDM model have led some researchers 
to propose alternative forms of dark matter (e.g., warm or self-interacting) 
to reduce the central masses of satellites \citep{Maccio2010, Vogelsberger2012, Lovell2012, 
Anderhalden2013, Shao2013, Polisensky2013}.  However, the highest resolution 
simulations available to date to study the internal properties of 
satellites include only the dark matter (DM) component of galaxies, 
neglecting the effects of baryons \citep[e.g.,][]{Diemand2007, 
Springel2008, Boylan-kolchin2012}. 

\citet{Zolotov2012} recently examined how baryons impact the DM structure 
in satellites around a MW-mass galaxy.  They demonstrate that supernova (SN) 
feedback reduces the central DM densities of satellites with $M_{\star} 
\gtrsim 10^7 M_{\odot}$ before infall \citep[see also][]{Governato2012}. 
However, SN feedback alone is not enough to bring the observed densities of 
satellites in line with observations.  After infall, the presence of a 
baryonic disk in the host galaxy increases the mass loss rate for all satellites 
via tidal stripping \citep{Zolotov2012, Arraki2013}. This tidal effect is 
particularly strong for those satellites that enter with cored DM halos 
\citep{Penarrubia2010}, further increasing the discrepancy in the central 
masses predicted by DM+baryon and DM-only simulations.  Previous studies have 
invoked tidal stripping after infall to reduce the DM masses of satellite galaxies 
\citep{Gnedin1999, Taylor2001, Hayashi2003, Kazantzidis2004, Kravtsov2004, 
Read2006b, Sales2007, Munoz2008, D'Onghia2010, Choi2009, Collins2011}.  
The additional central mass 
present in the parent halo of a baryonic run (due to the fact that baryons can 
cool to the center, unlike DM) lead to an enhanced tidal force not found in 
DM-only simulations, leading to enhanced stripping.  These earlier works 
examined the evolution in the densities of satellites using idealized models, 
because cosmological simulations were unable to achieve similar resolutions 
until recently.  

The results of these idealized models can be used to parametrize the stripping 
of mass based on the orbital history of a satellite.  However, even after 
considering the additional tidal stripping that should occur in a cosmologically 
motivated population of DM-only satellites, studies still could not reproduce a 
$z=0$ satellite population that matches the observed MW satellite kinematics.  
Not enough mass is stripped from the most massive satellites to bring them in 
line with the kinematics observed in the most luminous satellites of the 
MW \citep{Read2006b, D'Onghia2010}.  \citet{Read2006b} concluded that the most 
massive satellites would need to have central density slopes shallower than NFW 
models to undergo enough stripping to make the theoretical and observational masses 
consistent.  This is because subhalos become more prone to tidal forces as their 
density slopes become less steep \citep{Hayashi2003, Kazantzidis2004, 
Penarrubia2010}.

\citet{Zolotov2012} are the first to use cosmological simulations to probe the 
combined reduction in mass from both DM core creation and tidal stripping in the 
satellite population around a MW-mass galaxy.  In this paper, we use a complementary 
sample of satellites to extend the analysis presented in \citet{Zolotov2012} and 
further explore the observational consequences of the model.  We show that this 
model can, for the first time in fully cosmological simulations of MW-mass galaxies, 
reproduce the observed kinematics of dwarf Spheroidal (dSph) satellites in L$^*$ 
galaxies at $z=0$.  We use our results to interpret the observed trend and scatter 
in the dynamics of the MW and M31 dSphs.  We also demonstrate that the inclusion of 
baryonic physics leads to destruction of a number of luminous satellites that 
otherwise survive in DM-only simulations, and make new predictions for the 
surviving subhalo mass and luminosity functions of L$^*$ galaxies.  Finally, we 
compare our simulated dSphs to simulated field dwarfs of the same luminosity, and 
make predictions for observable differences in the two populations.

\section{The Simulations}

The simulations that we use here are drawn from \citet[][hereafter Z12]{Zolotov2012}, 
and we refer the reader to that paper for full details of the simulations.  Briefly, 
two halos with virial masses of $7\times10^{11} M_{\odot}$ and $8\times10^{11} M_{\odot}$ 
at $z=0$ were run with and without baryons.  The DM-only simulations were run with 
{\sc pkdgrav} \citep{Stadel2001}, while the simulations with baryons were run with 
{\sc pkdgrav}'s N-Body $+$ Smoothed Particle Hydrodynamics (SPH) successor, 
{\sc Gasoline} \citep{Wadsley2004}.  In the SPH runs, gas can cool via primordial 
and metal lines following \citet{Shen2010}.  A prescription for creation and 
destruction of H$_2$ is implemented following \citet{Christensen2012}, and 
star formation is tied directly to the presence of H$_2$ gas. 
A uniform UV background turns on at $z=9$, mimicking reionization 
\citep{Haardt2001}.  Supernovae deposit 10$^{51}$ ergs of thermal energy 
into the surrounding gas following the ``blastwave'' scheme described in 
\citet{Stinson2006}.  

\subsection{Satellite Sample}

In Z12, the satellite sample was identified using the SPH simulations.  That 
is, the sample included only satellites in the luminosity range of the ``classical'' 
MW dSphs, with $M_V$ brighter than $-8$, and within the virial radius of the main 
SPH halo, i.e., within $\sim$240 kpc of the center of the parent halo.  We then 
identified the matching counterparts of these surviving, luminous subhalos in the 
DM-only run.  In this work we take an alternative approach, and instead identify 
our sample based first on the surviving subhalos of the DM-only simulations, and 
then search for their matching counterparts in the SPH runs.  In addition, since 
most of the classical dSphs could be observed even beyond 240 kpc, we extend our 
search beyond the defined virial radius.  In this paper, we have chosen to include 
all halos with $v_{max} > 15$ km/s at $z=0$ that are within 300 kpc of the main halos.  

This new satellite selection yields 62 DM-only satellites combined between the 
two MW-mass halos.  This is more 
satellites than were presented in Z12, for three reasons.  
First, extending the search to 300 kpc nearly doubles the number of satellites in 
one of our halos, though adds only two satellites in the other.
Second, there are more satellites in the DM-only run than the SPH run at $z=0$.  This 
is due to baryonic processes, which more efficiently destroy satellites in the SPH run.  
As we'll show in Section~\ref{section4}, this destruction is most effective for 
satellites with small pericentric radii.  The primary destruction mechanisms are 
tidal heating and disk shocking, in which the SPH satellite passes very close to  
or through the parent 
galaxy's disk, which does not exist in the DM-only run.  Our two MW-mass halos 
have a combined total of eight satellites that survive in the DM-only runs, but 
which are destroyed in the SPH runs by $z=0$ (six of which are brighter than 
$M_V = -8$). Third, the luminous satellites selected in Z12 had DM-only 
counterparts that generally had $v_{max}$ at infall, $v_{@infall}$, greater 
than 25 km/s.  Our new satellite selection introduces lower mass halos.

The primary benefit of our new satellite selection is that it allows us to 
highlight in detail the discrepancies that occur between a DM-only run and a 
baryonic run due to the inclusion of baryonic physics.  Not only can we explore 
the source of the discrepancies, but demonstrate the erroneous conclusions that 
would result if one tries to use DM-only simulations to match observational 
results.

\subsection{Baryonic Evolution with Mass}

\begin{figure*}
\plotone{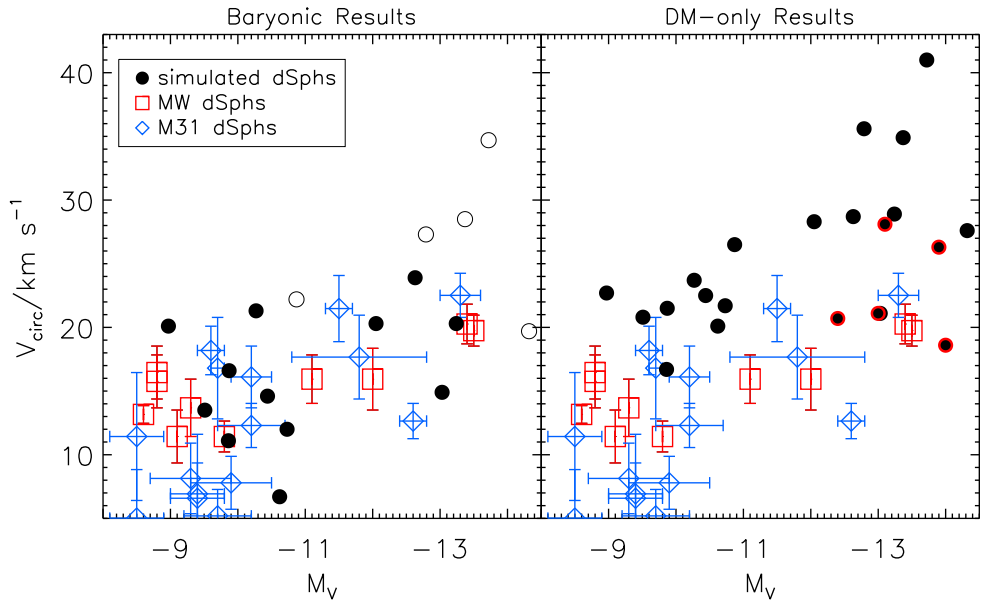}
\caption{Observational data from the MW and M31 satellites compared to the 
satellite populations of two simulated MW-mass hosts.
{\it Left:} $v_{max}$ at $z=0$ for the SPH satellites as a function 
of their $V$-band magnitude. The solid circles are gas-free at $z=0$, and 
are true dSph analogs.  The empty circles are SPH satellites that still 
contain gas.  The empty red squares and empty blue diamonds are $v_c$ at the 
half light radii for MW and M31 dSphs, respectively. 
{\it Right:} $v_{max}$ at $z=0$ for the satellites in the DM-only runs as 
a function of the $V$-band magnitude or their matched SPH counterpart. The
baryonic simulated dSphs have the same range of magnitudes and velocities 
as the observed dSphs, while the high $v_{max}$ values of the DM-only 
simulations are inconsistent with the observations. The circles with red rings
indicate satellites for which the SPH counterpart has been completely tidally 
destroyed.}
\label{z0}
\end{figure*}

Our concern in this paper is with luminous satellites bright enough to be 
considered classical dSphs, with $M_V$ brighter than $-8$.  These satellites 
are bright enough that their detection should currently be nearly complete 
out to 300 kpc in the MW \citep{Tollerud2008, Walsh2009}.  Z12 showed that 
the $z=0$ satellite populations of both simulated galaxies are in good 
agreement with the MW and M31 satellite luminosity functions, from 
$-8 < M_V < -15$.  Despite the inclusion of more satellites when searching 
out to 300 kpc, only one satellite in each of the two MW-mass simulations is 
added that is bright enough to be considered a classical dSph, and thus does 
not significantly alter the luminosity functions.  The remaining satellites 
that have now been included from beyond the virial radius would be too faint 
to identify given their distances and current observational limits.  As we 
discuss in the next paragraph, the lower mass halos that we've added are fainter than the 
classical dSphs, and therefore do not alter the luminosity functions.

The circular velocities of classical dSphs in the MW and M31 are typically below 
20 km/s, suggesting that our subhalo sample with $v_{max}$ down to 15 km/s might 
host luminous dSphs. However, we have found that our lowest mass halos do not 
form enough stars to be considered a classical dwarf.  We investigate this further 
by identifying $v_{@infall}$ for all of our halos.  Assuming
that $v_{@infall}$ corresponds to the highest mass that the subhalo had, prior to 
being tidally stripped after accretion onto the main halo, it should be the best 
predictor of a subhalo's luminosity \citep[e.g.,][]{Kravtsov2010, Rashkov2012, 
Hearin2013}.  
Indeed, Z12 found a tight correlation between 
$v_{@infall}$ and the stellar mass of the subhalos in these simulations.

We use AHF \citep{Knollmann2009,Gill2004} to identify halos and subhalos at every 
output step, and trace back the most massive progenitor of each satellite.  We 
find that the faintest halos are also the lowest mass at infall.  All of our halos 
with $v_{@infall} < 20$ km/s have stellar masses below 10$^5$ M$_{\odot}$, and 
$M_V$ fainter than $-7$. 
All of our SPH satellites brighter than $M_V = -8$ have $v_{@infall} > 20$ km/s.
However, there is some stochasticity in halos with $20 < v_{@infall} < 25$km/s.
In this mass range, some dwarfs are bright enough to be considered classical 
dSphs, but some are too faint.  

The nature of the star formation in the subhalos varies with mass.  All of 
the more massive, luminous satellites with $v_{@infall} > 25$ km/s are
able to retain gas until infall, allowing their star formation histories 
(SFHs) to extend at least until infall. On the other hand, halos with 
$v_{@infall} \lesssim 25$km/s typically lose their gas prior to accretion onto 
the parent halo.  Heating from the uniform UV background, combined with 
early star formation and SN feedback, removes a substantial amount 
of gas from these low mass halos, leaving them gas poor.  These halos 
lose $10-90$ times more mass in gas than stars formed by the time of their 
accretion.  The remaining gas has low surface densities and is inefficient at 
producing stars.  The low level of star formation that can occur prior to infall 
is relatively constant rather than stochastic.  

Halos with $v_{@infall} > 25$km/s are able to retain more of their gas for longer, 
allowing them to have extended SFHs \citep[in line with observational data, 
see][]{Grebel2004, Dellenbusch2008, Weisz2011}. In halos with $v_{@infall} > 30$km/s,
the SFHs tend to be episodic and bursty, unlike the lower mass satellites. With 
their deeper potential wells, the more massive halos achieve an initially higher 
star formation rate (SFR) than the lower mass halos as gas cools in the central 
galaxy.  The subsequent SN feedback following a burst of star formation heats the 
surrounding gas, shutting off star formation for a period of time until the gas 
can again cool and continue with another burst.  The bursty SFHs in the most 
massive halos lead to DM core creation prior to infall \citep[see 
also][]{Read2005, Maccio2012, Pontzen2012, Teyssier2012, Penarrubia2012, 
Garrison-kimmel2013}.\footnote[1]{Note that some authors have suggested that 
processes such as dynamical friction or angular momentum transfer are also 
responsible for DM core creation \citep[e.g.,][]{ElZant2001, 
Tonini2006, Delpopolo2009}.}

Many subhalos lose most of their gas after infall and are gas-free by 
$z=0$.\footnote[2]{Satellites that retain gas until $z=0$ may be artificially 
gas-rich, due to inefficient stripping of gas in this implementation of SPH 
\citep{Agertz2007}.}  
While some subhalos lose their gas nearly instantly at infall, some are capable of 
retaining their gas for an extended time, and even having a low level of star formation.  
However, the SFRs become strongly suppressed after infall, and no subhalo continues 
to undergo the bursty star formation that contributes to DM core creation.

\section{Interpreting Observed Dwarf Spheroidals}

Below we discuss the observational properties of the Milky Way and M31 
dSphs in light of results that show that baryonic effects can reduce the 
central DM masses of luminous satellites that infall onto a $\sim L^*$ galaxy. 

\subsection{Satellite Kinematics at Redshift 0}

Figure~\ref{z0} shows $v_{max}$ of the simulated satellites as a function of 
$M_V$. The simulated SPH satellites are shown in the left panel, while the 
DM-only satellites are shown in the right panel.  The $v_{max}$ values for the 
simulated gas-free satellites in the SPH run span the range of 6-24 km/s, 
consistent with $v_c$ results for MW and M31 dSphs (red and blue symbols, 
respectively). In contrast, all but 2 of the DM-only counterparts 
have $v_{max} >$ 20km/s, grossly inconsistent with the MW and M31 dSph data. 

The filled circles in the left panel of Figure~\ref{z0} show the gas-free SPH 
satellites (dSph analogs).  Satellites that still retain gas are shown for 
completeness as empty circles. Plotted with 
the simulation data are the observed circular velocities at the half light radii 
for dSphs with $M_V$ brighter than $-8$ in both the MW and M31.  The MW data 
is drawn from the compilation of \citet{McConnachie2012}.  The M31 data is 
taken from table 5 in \citet{Collins2013}, where overlapping results with 
\citet{Tollerud2012} are weighted based on the quality of the observations.  
For the observational data, $v_c (r_{1/2})= \sqrt{3\sigma^2}$, 
where $\sigma$ is the line of sight stellar velocity dispersion, and $r_{1/2}$ 
is the half-light radius of the dSph.  \citet{Walker2009} and \citet{Wolf2010} 
have shown that assumptions about isotropy are minimized at $r_{1/2}$, making  
the masses determined at $r_{1/2}$ the most robust. Because most dwarfs show 
flat $\sigma$ profiles \citep{Walker2009}, it is assumed that $v_c (r_{1/2})$ 
is roughly equal to $v_{max}$.  In any case, the half-light radii are 
$\lesssim$ 1 kpc for all of the observed dSphs.  Our measurements for the 
simulated satellites are upper limits to the values derived at typical 
half light radii, as the rotation curves of these satellites either peak or 
continue to rise at 1 kpc.  

\begin{figure}
\includegraphics[angle=0,width=0.5\textwidth]{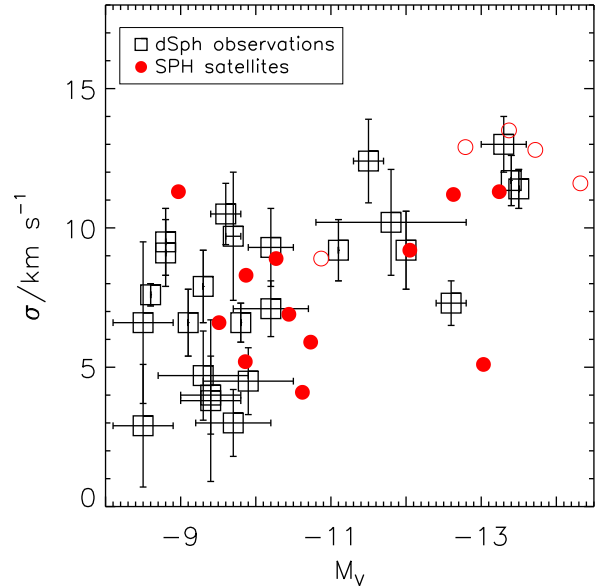}
\caption{Stellar line-of-sight velocity dispersions, $\sigma$, in satellites 
measured at the half light radius, as a function of $V$-band magnitude.  Data 
for MW and M31 dSphs are now shown collectively as empty squares, and the 
simulated satellites are in red.  As in 
Figure~\ref{z0}, the solid circles are gas-free SPH satellites at $z=0$, and 
are true dSph analogs.  The empty circles are SPH satellites that 
still contain gas.  This plot is similar to the left panel of Figure~\ref{z0}, 
but is a more robust test of the SPH simulations since it directly compares to 
the observational data. }
\label{sigma}
\end{figure}

Since the line of sight stellar velocity dispersions are the most 
direct comparison to observations, we show in Figure~\ref{sigma} a comparison 
of our simulated stellar dispersions to the observations. The same simulated 
SPH satellites shown in the left panel of Figure~\ref{z0} are shown in 
Figure~\ref{sigma}.  We have chosen a random line of sight. 
Similar to the observational data, we found that the stellar dispersions are 
relatively constant with radius.  Hence, in Figure~\ref{sigma} we plot the 
value of the dispersion measured at the half light radius, just as in the 
observations.  We find that the dispersions are in good agreement with the 
range of observed dispersions. This demonstrates that, although Figure~\ref{z0} 
is an indirect comparison of the mass enclosed between the observations and 
simulations, a more direct comparison measured at similar radii also confirms 
that the simulations with baryons are able to reproduce the observational results. 

The right panel of Figure~\ref{z0} shows the satellites in the DM-only 
simulation compared to the observational data.  Each DM-only subhalo is
assigned the $V$-band magnitude of its matched SPH satellite counterpart. 
As we discuss in Section~\ref{section4}, some of the DM-only subhalos survive 
to $z=0$, but their counterparts in the SPH run are destroyed.  These subhalos 
have been indicated by red rings surrounding 
the data points. For those subhalos, we assign the luminosity of the matched 
SPH subhalo at the time it is accreted to the parent halo, well prior to its 
disruption.  All of these destroyed satellites are brighter than $M_V < -12$, 
so there are many more bright DM-only subhalos than in the SPH run.

In Figures~\ref{z0} and \ref{sigma}, we restrict ourselves to examining the sample 
of halos that host classical, luminous dSphs. This neglects just over half of the 
subhalos that we have identified with $v_{max}(z=0) > 15$km/s and within 300 kpc, 
because they are fainter than $M_V = -8$ where the observational data is 
incomplete \citep{Tollerud2008, Walsh2009}.  
At the bright end, we also wish to ignore those satellites that are Magellanic 
Cloud analogs.  \citet{Boylan-kolchin2012} assign Magellanic status to those 
halos that had $v_{@infall} > 60$km/s and $v_{max}(z=0) > 40$km/s.  This removes 
one subhalo from our sample.  However, two additional satellites (one in each 
halo) have $v_{@infall} > 40$km/s, but were accreted less than 25Myr prior to 
$z=0$.  We remove these from the sample as well, as they are clearly dwarf 
Irregular (dIrr) analogs rather than dSphs.

Previous studies using DM-only simulations have found that subhalos simulated 
in a CDM-context are significantly more dense than observations of the MW dSphs 
suggest \citep{Boylan-kolchin2011, Boylan-kolchin2012, Wolf2012, Hayashi2012, 
Rodriguez-Puebla2013}.  However, when baryonic effects are modeled 
self-consistently, as they are here, the kinematics of simulated satellites 
are well-matched to the observed kinematics of dSph satellites.

\begin{figure}
\includegraphics[angle=0,width=0.5\textwidth]{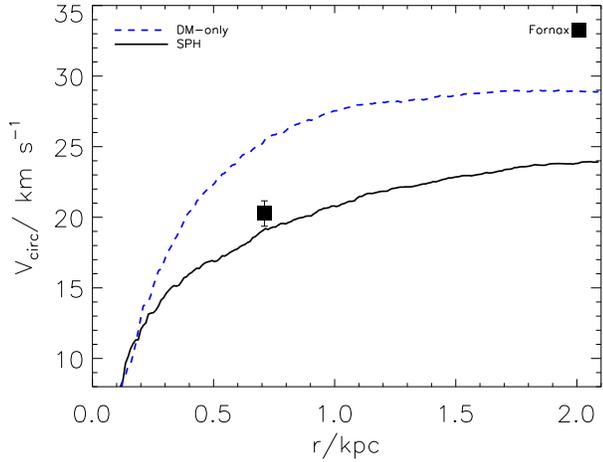}
\caption{The $z=0$ rotation curves of a simulated satellite  
and its DM-only counterpart. The $v_c$ for Fornax is over-plotted, 
based on the data in \citet{Walker2009}. The combination of 
SN feedback (before infall) and tidal stripping (after infall) substantially 
lower the $v_c$ of the SPH satellite by $z=0$, and is in good agreement with 
the observed $v_c$ of Fornax. }
\label{fornax}
\end{figure}

Figure~\ref{z0} demonstrates that DM-only simulations over-predict the central 
DM masses of satellite galaxies, and explains why results from such simulations 
yield satellite populations that are inconsistent with the kinematics of the 
MW dSphs.  An explicit example of the difference in kinematics for a bright 
satellites produced in a DM-only context compared with the same satellite in 
a baryonic simulation is shown in Figure~\ref{fornax}.
The total rotation curve (DM and baryonic mass) for an SPH simulated subhalo is 
shown in Figure~\ref{fornax}, which closely matches the derived $v_c$ at the half 
light radius of Fornax, along with the rotation curve for the same subhalo in a 
DM-only run without baryons. The DM-only satellite clearly over-predicts the 
central mass of this satellite.  For almost all of our satellites, the 
{\it DM-only runs produce satellites with $2-4$ times more mass in the central 
1 kpc than their SPH counterparts}.  

It is common when making predictions for the properties of subhalos within the 
CDM model to associate the most massive DM-only subhalos at infall with the most 
luminous satellites at $z=0$ \citep[e.g.,][]{Koposov2009, Boylan-kolchin2011, 
Boylan-kolchin2012}, i.e., an abundance matching technique.  Indeed, the most 
massive subhalos in the DM-only runs in this work are matched to the most luminous 
satellites in the SPH runs at infall (as shown in Z12).  However, SPH satellites 
experience evolution that DM-only runs do not account for.  It is therefore 
incorrect to assume that the central masses of DM-only subhalos should predict 
the observed masses in the inner regions of luminous subhalos.

\subsection{The Scatter in $v_c$}

A long-standing puzzle in the observed data shown in Figure~\ref{z0} is that 
some of the bright MW dSphs with $M_V \sim -13$ have similar $v_c$ values to 
some of the much fainter galaxies with $M_V \sim -8$.  In the field, galaxies 
display a clear trend of increasing stellar mass with halo mass 
\citep[e.g.,][]{Behroozi2010, Moster2012, Leauthaud2012, Behroozi2013}, yet 
satellites that span two orders of magnitude in luminosity seem to show no trend 
in their central $v_c$ with mass. The comparable velocities of Draco ($M_V = -8.8$), 
and Fornax ($M_V = -13.4$), for example, have led \citet{Penarrubia2008} to 
conclude that Draco formed in a halo 5 times more massive than Fornax, despite 
being 70 times fainter.  

\begin{figure}
\includegraphics[angle=0,width=0.5\textwidth]{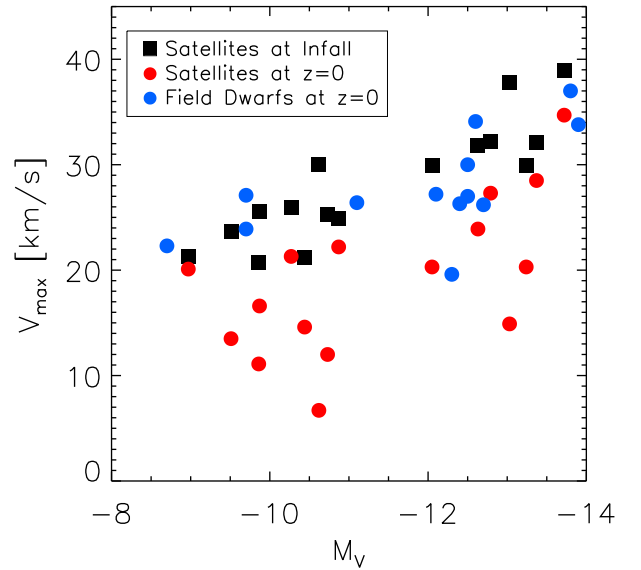}
\caption{The $v_{max}$ values of simulated dwarfs. The black squares show the simulated 
satellites at the time of their accretion.  The red circles show the satellites' 
$v_{max}$ at $z=0$ (as in Figure~\ref{z0}).  The blue circles show $v_{max}$ 
at $z=0$ for simulated field dwarfs in the same luminosity range as the classical 
dSphs (discussed further in Section 5).  }
\label{evol}
\end{figure}

The simulated SPH satellites shown in Figure~\ref{z0} follow a similar trend to 
the observed MW and M31 dSphs; at $z=0$ simulated satellites with $M_V = -9$ can have 
comparable kinematics to satellites with $M_V = -13$.  At infall, however, SPH 
satellites follow an increasing trend of stellar mass with halo mass, with 
M$_* \propto$ M$_{vir}^2$.  This is shown in Figure~\ref{evol}. The 
trend seen for satellites at infall (black squares) and in the field at $z=0$ 
(blue circles; discussed further in Section 5) is erased in the subsequent 
tidal evolution of the satellites, resulting in the scatter observed 
at $z=0$ (red circles).   

Since both SPH and DM-only satellites undergo tidal stripping, one might expect 
that SPH and DM-only satellites would display a similar scatter in $v_c$ at 
$z=0$. However, when DM-only satellites are matched to the stellar mass of their 
SPH counterparts, DM-only satellites still show a trend of increasing $v_{max}$ 
with luminosity at $z=0$, unlike satellites simulated with baryons.  This can be 
seen in the right panel of Figure~\ref{z0}, particularly when one neglects those 
satellites that have destroyed SPH counterparts (with red rings).  The DM-only 
satellites with destroyed SPH counterparts do undergo much more tidal stripping
than other subhalos, but in the presence of a disk this should have led to their 
total disruption.  When the DM-only satellites with surviving SPH counterparts 
are considered, there is a trend of increasing $v_{max}$ with luminosity.

Two effects lead to the divergence in the scatter of $v_c$ between SPH and 
DM-only satellites. First, the presence of a baryonic disk\footnote{Note that the 
baryons need not be in a disk.  Any concentration of baryons, even in a spherical 
bulge, still leads to more tidal stripping than in the DM-only case \citep{Chang2013}.}  
in the host galaxy enhances the rate of mass lost at each pericentric passage for 
all SPH satellites.  In essence, because the baryonic simulations have a component 
(gas) that can cool to the center of the parent halo, there is more mass sitting at 
the center of the parent halo, which leads to stronger tidal forces. Using the same 
set of simulations as this paper, we showed 
in Z12 (see Figure 6 in Z12) that SPH satellites undergo a larger reduction in their 
central circular velocity after infall than DM-only satellites. In that work, we showed 
that this effect is partially due to the baryonic disk in the SPH simulated host, and  
is increasingly significant with increasing time after infall and for satellites with 
smaller pericenters \citep[see also][]{Penarrubia2010}.
The variation in infall 
times and orbital pericenters results in the large spread of $v_c$ values for 
simulated SPH satellites fainter than $M_V \sim -12$, similar to the observational 
data in Figure 1. Because DM cannot cool to form a disk, the primary host in the 
DM-only simulation has no central disk component, and the low-mass DM-only subhalos 
that come near the central regions of the host do not experience enhanced tidal 
stripping.  Hence, they do not reproduce 
the observed scatter in $v_c$ at low masses (or equivalently, at low $M_V$). 

The second effect at play in creating the scatter in $v_c$ for SPH satellites is 
that the influence of a baryonic disk is especially strong for satellites with 
shallow DM density profiles, like the SPH satellites in our sample brighter than 
$M_V \sim -12$.  A shallower density profile means that the halo has a lower binding 
energy, which exacerbates the amount of central mass lost in the most luminous 
subhalos due to tidal stripping. The satellites fainter than $M_V = -12$ have 
not had enough star formation to 
create shallow DM density profiles or large DM cores \citep{Penarrubia2012}, 
so the reduction in mass is strictly due to tidal stripping.  For the brighter 
satellites, tidal stripping effects are even more severe due to the presence of a 
shallow DM density profile.  The reduction in DM density due to DM core creation 
alone is insufficient to reduce the central densities of the brighter satellites 
to the low $v_c$ values observed \citep[see also][]{Garrison-kimmel2013}.
However, the presence of a DM core in the satellite leads to enhanced tidal 
stripping that contributes to lowering the central $v_c$ of these satellites to 
under 25 km/s.  
The result of these processes is that satellites across the luminosity range 
appear to have similar masses at $z=0$. Without the effect of the disk, and with 
their steep density profiles, DM-only satellites experience overall less mass 
loss for a given orbit, and therefore tidal effects do not introduce a large 
scatter in their $v_c$. 

\section{The Role of Tides}
\label{section4}

\begin{figure}
\includegraphics[angle=0,width=0.5\textwidth]{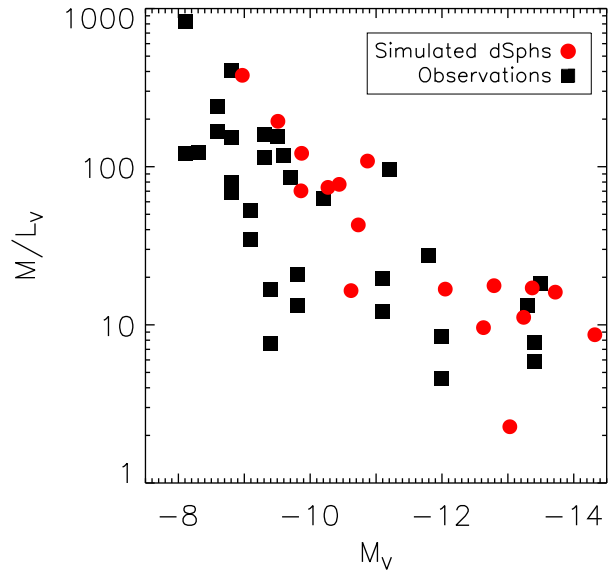}
\caption{The mass-to-light ratio, M/L, within the half light radius 
for the simulated SPH satellites (red circles) as a function of 
$V$-band magnitude.  Data from dSphs in the MW \citep{McConnachie2012} 
and M31 \citep{Tollerud2012, Collins2013, Tollerud2013} are shown as 
black squares.  The simulations match the observed trend of increasing 
M/L ratio with decreasing luminosity.}
\label{m2l}
\end{figure}

It is well established that dSphs are DM dominated galaxies \citep[e.g.,][]
{Mateo1998,Simon2007,McConnachie2012}.  Given that the baryonic processes 
described here all lead to decreasing the DM mass in a satellite before any 
stars are stripped, a natural question is whether these processes can still 
result in galaxies that are as DM dominated as the observed dSphs in the 
Local Group. 

Figure~\ref{m2l} shows the mass-to-light (M/L$_V$) ratios within the half light 
radius for the baryonic simulations as a function of $M_V$ (red circles).  The 
solid black squares are observed dSphs.  The MW data is taken from tables 3 
and 4 in \citet{McConnachie2012}, while the Andromeda data is compiled from 
\citet{Tollerud2012}, \citet{Collins2013}, and \citet{Tollerud2013}.  The 
simulations match the observed trend of increasing M/L ratio with decreasing 
luminosity.

The driving property behind this M/L trend is the relation of stellar 
mass to halo mass for the simulations, M$_{star} \propto$ M$_{halo}^2$.  As 
discussed above, while some subhalos lose substantial amounts of DM after 
infall, most lose only a factor of $2-4$ from their central regions.  This small 
change is not enough to significantly alter the M/L ratio.  Hence, the trend 
in M/L ratio is driven by the luminosity as a function of halo mass\footnote[3]{For 
stellar masses above 10$^6$ $M_{\odot}$, this relation is in agreement with 
commonly adopted relations in the literature, \citep[e.g.,][]{Koposov2009,
Kravtsov2010, Rashkov2012,Sales2013}.  At lower stellar masses, these same 
authors find a slightly steeper relation, with M$_{star}$ scaling as 
M$_{halo}^{2.5-3}$.  This difference between our simulation relation and derived 
relations has little to no impact on the results in this paper, but may explain 
the tendency for the fainter simulated satellites to lie systematically to 
slightly higher $M_V$ than observations in Figure~\ref{m2l}.} instead of mass 
loss.

\subsection{Orbital Dependence}

\begin{figure}
\includegraphics[angle=0,width=0.5\textwidth]{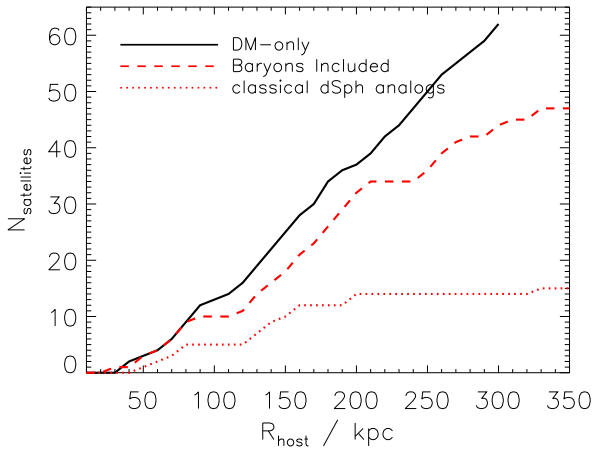}
\caption{The cumulative distribution of radial distance from the halo center
for the subhalos.  The solid black line shows the results for the DM-only 
runs, while the dashed red line is for the baryonic runs.  Destruction of 
subhalos in the presence of a baryonic disk leads to a depletion of the 
SPH satellites compared to the DM-only runs.}
\label{rz0}
\end{figure}

Z12 explored the difference in mass tidally stripped after infall as a 
function of pericentric distance, $R_{peri}$.  Their figure 6 shows that 
significant discrepancies between DM-only and baryonic runs can occur for 
$R_{peri} \lesssim 40$ kpc due to the stronger tidal forces present when 
baryons cool to the center of the halo.  
We have verified that the $R_{peri}$ values between matched counterparts in 
the DM-only and baryonic simulations do not show any global differences.  
For example, one might expect that the SPH satellites could have smaller 
$R_{peri}$ if adiabatic contraction occurred in the parent halo, but we find 
no evidence for this.  

In this section we discuss two effects that, in addition to tidal 
stripping, are likely to be important in understanding the evolution of 
satellites -- tidal heating and disk shocking. 
Within $\sim20$ kpc of the galaxy
center, tidal heating and disk shocking are the dominant processes that lead 
to mass loss from subhalos \citep[see][]{D'Onghia2010}, and were not discussed 
in Z12.  Tidal heating is efficient for eccentric orbits and can destroy 
satellites faster than tidal stripping \citep{Gnedin1999}.  Disk shocking 
occurs when the subhalos pass through the dense baryonic disk.  The rapidly 
varying gravitational field adds energy to the subhalos, altering the 
structure, and quickly leading to dissolution \citep{Ostriker1972}.

In our simulations, 8 of the original 62 DM-only subhalos do not have a 
surviving SPH counterpart at $z=0$.  Six of these subhalos have pericentric 
passages that take them within 30 kpc of the galaxy center, 
$16 < R_{peri}/kpc < 29$.  For comparison, the radial extent of the parent 
halos' disks are 
$7 < r/kpc < 10$ at $z=1$, and $15 < r/kpc < 20$ at $z=0$.  The other two 
disrupted subhalos are subject to the cosmological context of the simulations, 
encountering fly-bys with nearby halos that lead to their destruction prior 
to accretion onto the main halo.  The destruction of these two halos emphasizes 
that, while the main culprit that reduces the subhalo population is the disk 
of the parent halo, interactions with nearby galaxies can also deplete the 
subhalo population \citep[see also][]{Kravtsov2004}. 

The radial distribution of the subhalos at $z=0$ is shown in Figure~\ref{rz0}.
Figure~\ref{rz0} shows that there is only one surviving subhalo within 30 kpc of 
the galaxy center in the baryonic run at $z=0$, (and no bright, classical dSphs 
interior to 40 kpc, dotted red line), consistent with the results in both 
\citet{D'Onghia2010}, and \citet{Romano2010}.  This is also consistent with the 
bright satellite distribution in the Milky Way, where only Sagittarius is 
observed within 30 kpc, and is in the process of being disrupted.

The satellite distribution in the DM-only run and baryonic runs is very similar 
out to $\sim$80 kpc, and then begins to diverge (black solid line and red dashed 
line).  There is a tendency for the SPH satellites in the outer halo to reside 
at larger distances than their DM-only counterparts.  This is consistent with 
later accretion times in the SPH runs than the DM-only runs, as also seen by 
\citet{Schewtschenko2011}.  

All but one of the surviving DM-only 
satellites that have destroyed SPH counterparts are further than 80 kpc from 
the galaxy center at $z=0$.   In other words, most of the surviving DM-only 
subhalos with destroyed counterparts are found beyond $\sim$80 kpc at $z=0$, 
but were within 30 kpc at some point in their orbital history.  Thus, when 
considering the halos within a DM-only run that may have been completely 
destroyed by baryonic effects, the entire orbital history needs to be 
accounted for.  

In summary, the presence of a disk acts not only to reduce the overall masses 
of the SPH subhalos via tidal stripping, but it can also reduce the number 
of satellites overall.  We quantify these combined effects further in the next 
section.

\subsection{The Velocity and Luminosity Functions}

\begin{figure*}
\plottwo{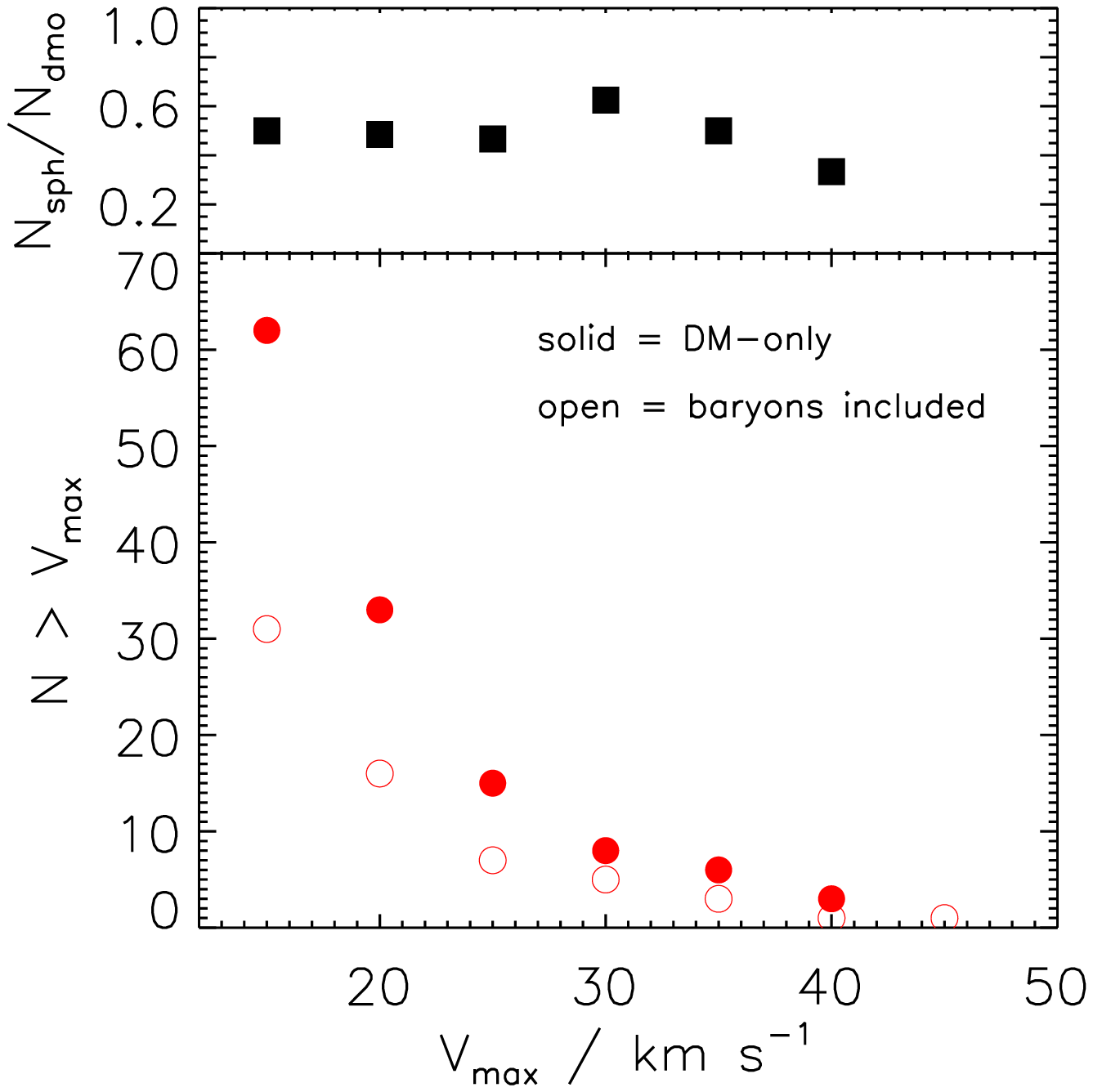}{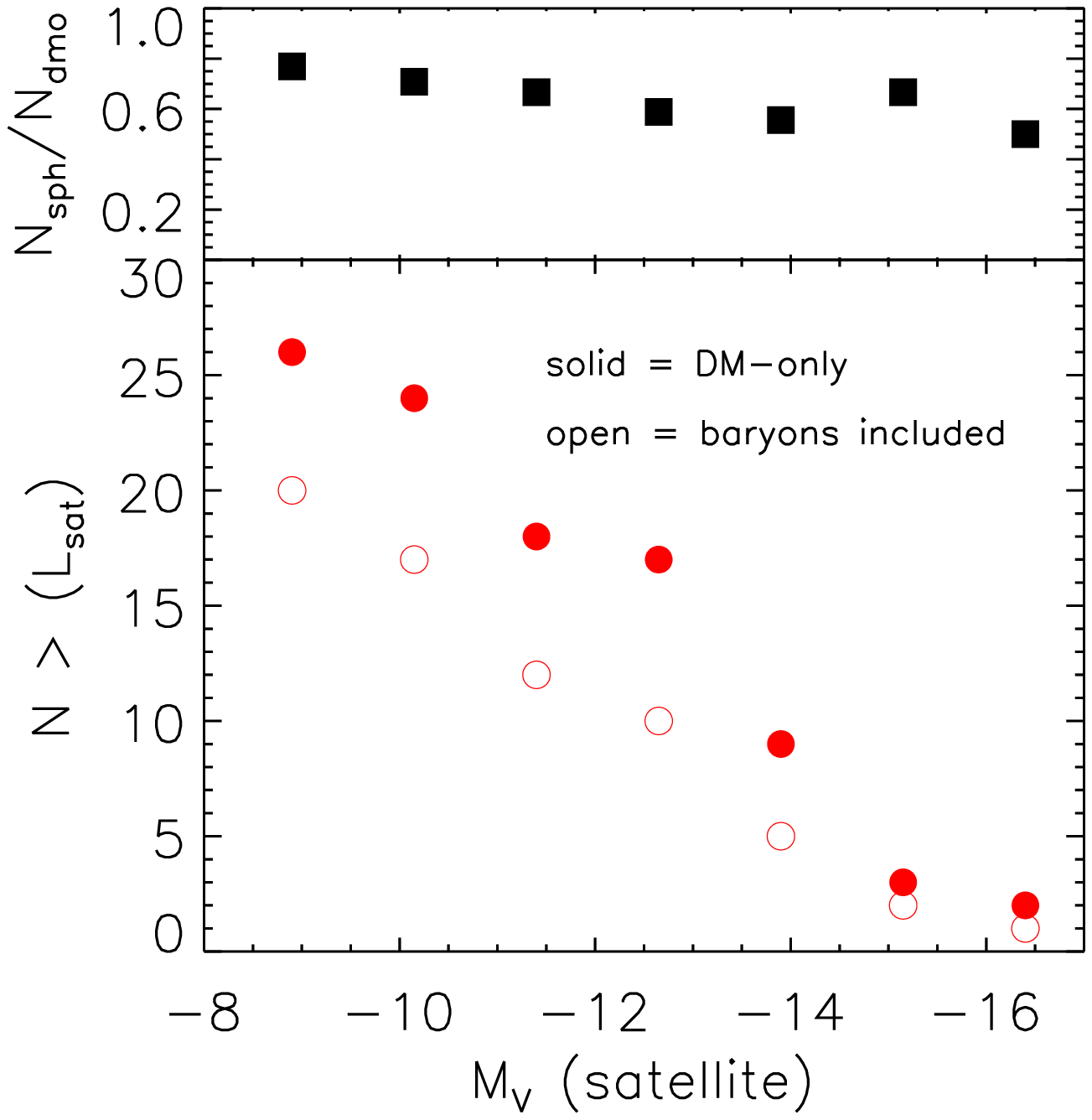}
\caption{The $z=0$ cumulative satellite velocity functions and luminosity 
functions.  Solid circles show the combined results for the two DM-only 
simulations, while the open circles show the combined results for the two SPH 
simulations.  {\it Left:} The $v_{max}$ function for the satellites.  All 62 
subhalos identified in the DM-only runs with $v_{max}$($z=0$) $>$ 15km/s and 
within 300 kpc of the main halos are shown here, compared to surviving SPH counterparts 
in the same velocity range.  The combined effects discussed in this paper (DM core 
creation, enhanced tidal stripping due to the presence of a disk, and tidal 
heating/disk shocking) all lead to a reduced number of subhalos at a given $v_{max}$ 
in the baryonic run compared to the DM-only runs.  The top plot shows that the number 
of subhalos is reduced by roughly a factor of two across the entire subhalo mass 
range we consider here.  
{\it Right:} The cumulative luminosity function for the satellites.  In this panel  
we use only the luminous satellites where observations are complete, $M_V$ brighter 
than -8.  We also include dIrr/Magellanic Cloud analogs.  The luminosity function 
is altered when satellites are completely destroyed in the baryonic runs.  The top 
plot shows that there are roughly 1/3 fewer luminous satellites in the baryonic runs 
compared to the DM-only runs.} 
\label{func}
\end{figure*}

Figure~\ref{func} shows the combined effects of mass loss and complete disruption on 
the resulting velocity and luminosity functions of the satellites.  Solid circles show 
the combined results for the two DM-only simulations, while the open circles show 
the combined results for the two SPH simulations.  The left hand panel shows the 
$v_{max}$ function.  In this panel we include all 62 of the subhalos identified in 
the DM-only runs that have $v_{max}$($z=0$) $>$ 15km/s and are within 300 kpc of the 
main halos.  

The $v_{max}$ function can differ quite dramatically between the SPH and 
DM-only satellites, due to multiple physical effects.  First, in a luminous subhalo 
($M_V$ brighter than $-12$), SN feedback can induce DM core creation.  DM 
is pushed out of roughly the central kpc \citep{Pontzen2012}.  This redistribution 
of the DM acts to lower $v_{max}$ in the baryonic run to $\sim$90\% of the value 
in their DM-only counterparts. 
Second, the presence of a disk leads to enhanced tidal stripping 
for {\it all} of the subhalos.  This process depends on infall 
time and orbital history for any given satellite, and therefore does not yield a 
smooth trend of mass loss as function of $v_{max}$, but is instead stochastic. 
Third, the number of subhalos at a given 
$v_{max}$ is reduced because some of the baryonic subhalos are destroyed entirely.
While this final effect removes halos entirely, the first 
two effects instead shuffle the $v_{max}$ distribution from that found in the 
DM-only runs.  This redistribution always acts to move subhalos to lower $v_{max}$, 
so that the cumulative number above a fixed $v_{max}$ is always lower in the 
baryonic runs compared to the DM-only runs.  The top plot in the left panel of 
Figure~\ref{func} shows that, at any given mass or $v_{max}$, there are roughly 
half as many subhalos in the baryonic run compared the the DM-only run \citep[see 
also][]{D'Onghia2010}.  Note that this reduction leads many of the SPH subhalos 
to have $v_{max} < 10$ km/s at $z=0$, and are not shown on this plot.

The right hand panel of Figure~\ref{func} shows the luminosity function of the 
satellites.  In this panel we neglect the faint halos ($M_V$ fainter than -8) 
where observational tests are incomplete, so there are fewer than half as many 
subhalos in the right panel as in the left panel.  However, unlike Figures~\ref{z0} 
and \ref{m2l} that exclude 3 bright dIrr/Magellanic Cloud analogs, we include the 
bright dIrrs here in order to demonstrate that discrepancies exist between 
the SPH and DM-only runs even up to the brightest, most massive satellites.
Unlike the $v_{max}$ function, where processes could remove DM mass and shift 
surviving subhalos to lower $v_{max}$, there is no shifting to lower luminosities 
that can occur.  In other words, tidal stripping can remove up to 90\% of the 
DM mass in a subhalo, but leave the tightly bound stellar content in place 
\citep{Munoz2008, Penarrubia2008, Libeskind2011, Chang2013}.  This alters the 
$v_{max}$ function, but not the luminosity function.  Instead, when tidal effects 
begin to finally disrupt the stellar content, the subhalo is soon thereafter 
completely destroyed.  Hence, the primary effect that alters the luminosity function 
of the subhalos is complete destruction.  Six luminous satellites, or roughly 1/3 
of the luminous satellites, are destroyed in the baryonic runs that survive in the 
DM-only runs (the two other destroyed satellites are fainter than $M_V = -8$).  
Hence, the top plot of the right panel in Figure~\ref{func} shows that the 
baryonic runs contain roughly 2/3 the satellites compared to the 
DM-only runs.

\section{Comparison to Field Dwarfs}

\begin{figure} 
\includegraphics[angle=0,width=0.5\textwidth]{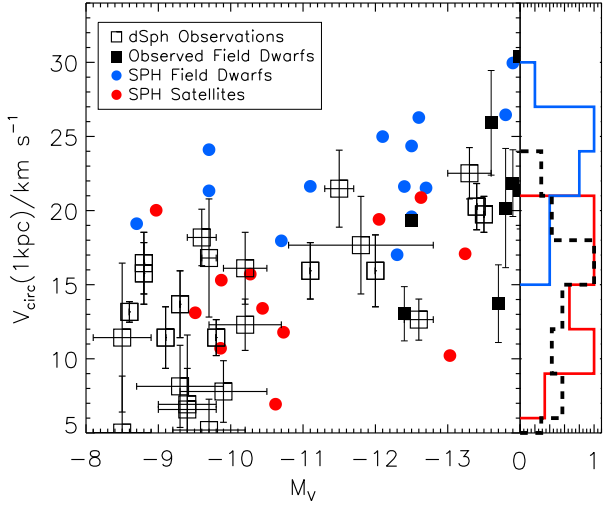}
\caption{The empty squares show observational data for the MW and M31 dSphs 
($v_c$ at the half light radii), as in Figure~\ref{z0}.  The remaining data 
points in this plot show $v_c$ at 1kpc.  Simulated dwarf {\it field} galaxies 
are shown by blue circles, and simulated satellites are shown by red circles.
Values for $v_c$ at 1kpc in observed dwarf field galaxies \citep[Oh, 
{\it private comm.};][]{Cote2000,McGaugh2007} are show by black squares.  
Histograms of the $v_c$ distributions are shown on the right of the plot.  
The blue line shows the $v_c$ for the simulated field dwarfs, the red line 
shows the $v_c$ for the simulated satellites, while the dashed black line 
shows the $v_c$ distribution for the observed dSphs for comparison.  
Although dwarf field galaxies may span the same magnitude range as 
the observed satellites, they systematically reside at higher $v_c$ than 
observed in dSphs.  Tidal stripping is required to remove enough mass in 
these luminous dwarfs in order to reach the range of observed $v_c$ in dSphs.}
\label{field}
\end{figure} 

Mass loss due to tidal effects is the essential component to reproduce the
range of observed $v_c$ in dSphs.  We demonstrate this in Figure~\ref{field}.
The observational data for the satellites (empty squares) shown in 
Figure~\ref{field} is derived again from dispersions measured at the half 
light radii, as in Figure~\ref{z0}.  We again plot our simulated satellites, 
but this time we show $v_c$(1kpc) rather than $v_{max}$ as before.  Also shown 
as solid squares are observed field galaxies, all measured at 1kpc,  
from {\sc Little THINGS} 
(Oh, {\it private comm.}), \citet[][for the Sagittarius dwarf Irregular 
galaxy]{Cote2000}, and \citet[][for DDO\,154]{McGaugh2007}.  The {\sc Little 
THINGS} data have been corrected for asymmetric drift, and include only galaxies 
for which a tilted ring analysis could be done.  We have used a larger sample of 
seven ``zoom-in'' simulations (including the two simulations from which our 
simulated satellite data is drawn) to identify isolated (non-satellite) dwarf 
galaxies in the same luminosity range of the classical dSphs (blue circles).  
Five simulations have the same resolution as the satellite results 
($\epsilon$ = 174pc), but two have even higher resolutions ($\epsilon$ = 87pc).

The values for the observed dwarf field galaxies in Figure~\ref{field} (solid 
black squares) represent $v_{rot}$ measured at $\sim$1 kpc.  Unlike the 
simulated dSphs sample, where $v_{max}$ is comparable to $v_c$ measured at 1 kpc 
($v_{\rm{1kpc}}$), $v_{max}$ of the simulated field dwarfs can occur at much larger 
radii ($3-11$ kpc in the luminosity range examined here, $-8 > M_V > -14$).  Since 
the half light radii of observed dSphs are typically 1 kpc or less, we measure 
$v_{\rm{1kpc}}$ for the field dwarfs for the most direct comparison.  
Figure~\ref{field} demonstrates that the mass enclosed in the field sample 
is systematically higher than that in observed dSphs.  While observed dSphs all 
have $v_c < 25$km/s, the field dwarfs {\it in the same luminosity range} instead have 
$15 < v_c < 30$km/s.  Using $v_{max}$ for the field dwarfs instead of $v_{\rm{1kpc}}$ 
would shift this range to 20-40 km/s (see Figure~\ref{evol}).  

In general, the observed $v_{rot}$ for field dwarfs at 1 kpc is in agreement 
with our simulated field galaxies.  Two of the {\sc Little THINGS} dwarfs lie 
below the range we predict, but we stress that the data is preliminary, and that 
these same set of simulations generally match the central densities of the THINGS 
and {\sc Little THINGS} sample at 500pc \citep{Oh2011,Governato2012}.  We also 
note that there is a strong luminosity bias, so that only galaxies brighter than 
$M_V = -12$ have been observed.  We leave it to future work to more closely 
evaluate whether our simulated field dwarf sample matches the velocity function 
of observed dwarfs \citep[e.g.,][]{Papastergis2011, Ferrero2012}.  For now, we 
predict the field sample should lie systematically to higher $v_c$ values than 
dSphs (or dwarf Ellipticals assuming this trend continues to brighter satellites).

Figure~\ref{evol} demonstrates that the $v_{max}$ of the satellites at the 
time of their accretion is in the range $20 < v_{max} < 40$km/s, comparable  
to the range of the simulated field dwarf galaxies at $z=0$ presented here.  Hence, 
the satellites start off with comparable masses to field dwarfs, but it is their 
subsequent mass loss through tidal stripping that results in the range of 
lower $v_c$ values that is actually observed.  In other words, while there 
exist field galaxies with $v_{\rm{1kpc}}$ comparable to observed dSphs (in the 
$15-25$ km/s range), {\it we predict that field galaxies in this luminosity range
should not span the full range of low $v_c$ that are observed in dSphs.}  
The exception to this rule might occur if field galaxies have previously 
undergone a fly-through with a more massive galaxy \citep{Sales2007b, Oman2013, 
MTeyssier2012}.  However, these galaxies should presumably be fewer in number 
than field galaxies that have avoided strong interactions, so that a population 
of satellite galaxies should show systematically lower central masses 
($v_{\rm{1kpc}}$) than field galaxies at a fixed luminosity. 

\begin{figure}
\includegraphics[angle=0,width=0.5\textwidth]{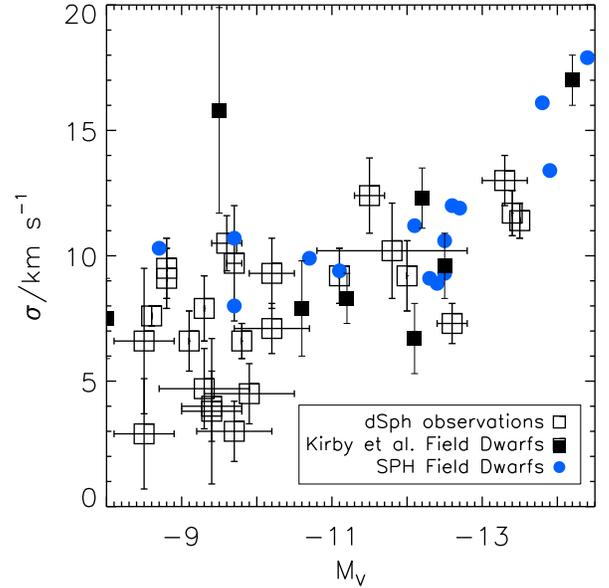}
\caption{Stellar line-of-sight velocity dispersions, $\sigma$, in dwarfs measured at 
the half light radius, as a function of $V$-band magnitude.  Data for MW and 
M31 dSphs are shown collectively as empty squares.  The solid squares are dispersion 
values measured in field dwarfs by \citet{Kirby2014}, who found that the dispersions 
of the field galaxies and the satellites show no significant offset.  Solid blue 
circles are simulated SPH field dwarfs, and are in agreement with the results 
of \citet{Kirby2014}.  The field dwarfs do not seem to have dispersions that 
span to the lowest values observed in the satellites ($\sigma <$ 5km/s).}
\label{sigmaf}
\end{figure}

A recent measurement of stellar velocity dispersion in \citet{Kirby2014} 
allows us to begin to test the tidal scenario.  \citet{Kirby2014} measured velocity 
dispersions within the half light radius for a number of field dwarfs in the same 
luminosity range as the dSphs.  They found that there is no significant difference 
in the dispersions of field and satellite dwarfs.  We compare our simulated field 
galaxies to the \citet{Kirby2014} sample in Figure~\ref{sigmaf}.  
Like the field observations, the dispersions of our simulated field dwarfs also 
show no significant offset from the dSph observations.  However, we note that, in 
both the observations and simulations, field galaxies never seem to span to the 
lowest $\sigma$ values observed in the satellites.  

\citet{Kirby2014} removed any rotational component from the kinematics, but found 
little rotation in their sample.  Likewise, our simulated field dwarfs show little 
net rotation at the small radii probed in Figures~\ref{field} and \ref{sigmaf}.  
At face value, this suggests that both the field and satellites 
galaxies are dispersion dominated, and have similar $v_c$ values (or at least that 
the field dwarfs do not show significantly larger $v_c$ values then the satellites).  
Figure~\ref{sigmaf} demonstrates that our simulated field dwarfs 
agree with this picture {\it when dispersions are measured within the half light 
radius of the galaxies}.  The half light radii of these dwarfs are generally much 
smaller than 1 kpc.  On the other hand, Figure~\ref{field} demonstrates that a 
clearer separation in $v_c$ begins to appear between the two samples as further 
radii are probed.  We predict that future measurements that can probe out to  
1 kpc or more will begin to see this offset that has been created by tidal 
stripping in the satellites. 

\begin{figure}
\includegraphics[angle=0,width=0.5\textwidth]{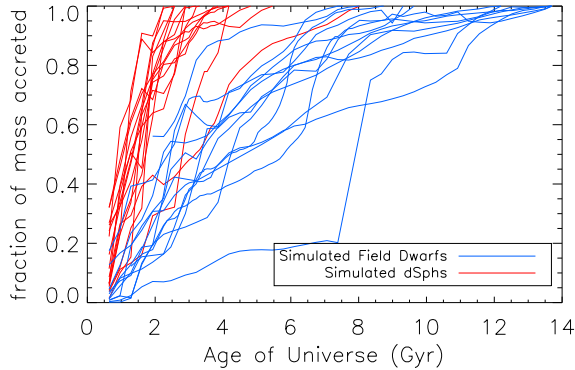}
\caption{The mass accretion histories of the simulated SPH satellites (red 
lines) and the simulated dwarf field sample (blue lines), both in the 
magnitude range $-14 < M_V < -8$.  The MAH of the 
satellites is truncated at the time of their infall onto their parent halo.
Satellites tend to be accreted at $z > 1$, meaning that they reach, at much 
higher $z$, similar $v_{max}$ values to the field dwarfs at $z=0$.  This 
also means that they reside in more concentrated halos than typical field 
dwarfs observed at $z=0$. } 
\label{mah}
\end{figure}

The fact that the observed satellites and observed field dwarfs have an 
overlapping range of $v_c$ values raises a finer point.  A naive 
conclusion might then be that satellites and field dwarfs at a fixed 
luminosity were born in similar halos.  However, this is not the case.  As 
was just stated, the $v_{max}$ range of the dSphs {\it at infall} is comparable 
to the $v_{max}$ range of field dwarfs {\it at z=0} (see Figure~\ref{evol}), 
and this is an important distinction.  The dSphs reached this $v_{max}$ at 
much higher redshift than the field sample.  

We demonstrate this in Figure~\ref{mah}, where we show the mass accretion 
histories (MAH) of our simulated satellites and simulated field dwarfs.  These 
simulated dwarfs all lie in the magnitude range $-14 < M_V < -8$, so the 
satellite sample excludes three dIrr/Magellanic Cloud analogs that tended to 
fall in at recent times.  The MAH of the satellites is truncated at the infall 
time onto the parent halo for clarity (the masses decline after infall, as we 
have emphasized in this paper).  Most satellites in our sample were accreted 
at $z > 1$. One of the consequences of the fact that satellites have an earlier 
MAH than field dwarfs is that satellites reside in much denser halos than typical 
field dwarfs at $z=0$. This is a natural consequence of galaxy formation in the 
CDM cosmological model, as galaxies that form at higher redshifts are more
concentrated \citep{Wechsler2002, Zhao2003}, and early forming subhalos are 
found preferentially closer to the center of MW-mass galaxies \citep{Diemand2005, 
Moore2006}. 

The more concentrated halos and deeper potential wells of the satellites compared 
to the field dwarfs could be manifest in the star formation histories (SFHs) of 
the two populations.  Assuming that SFHs are truncated or significantly lowered 
after infall of the 
satellites due to ram pressure stripping of their gas, the cumulative SFHs 
of the satellites should be biased to higher $z$ than field dwarfs, which 
exhibit continued SF all the way to $z=0$.  We verfied that this is the case; 
the simulated dSphs show an earlier peak in their cumulative SFHs than the 
simulated field dwarfs.  A plot of the cumulative SFH looks very similar to 
the MAH shown in Figure~\ref{mah}, so we do not show them here.  We will 
instead present them in a comparison to Local Group data in a future paper 
(Weisz et al., {\it in prep.}).  

Observationally, there is a hint that dSphs do have a faster rise in SF at 
early times compared to dIrr in the ANGST data \citep{Weisz2011}.  However, 
ANGST also contains the dynamically young group M81 \citep{Nichols2013}, and 
within errors the cumulative SFHs of the dSphs and dIrr are indistinguishable.  
New work comparing the cumulative SFHs of dSphs and dIrr in the Local Group
suggest a larger difference, with the SFHs of dSphs peaking closer to $z=1$ 
(Weisz et al., {\it in prep.}), though again the error bars are large. 
We also note that some 
MW dSphs show evidence for very recent SF, within the last $\sim$1Gyr 
\citep{Grebel2004, Dellenbusch2008}.  This is not excluded by the model, but 
depends strongly on infall time for any given halo, how long a satellite is 
able to retain gas after infall, and whether any re-accretion occurs 
\citep{deBoer2013}.  However, even if a few dSphs continue SF to low $z$, the 
population as a whole should show much earlier SF on average, according to 
our model, than a population of field dwarfs at similar luminosities.

Thus, although field dwarf masses may be similar to dSphs at a fixed luminosity, 
this does not mean that tidal processes in the dSphs are unimportant.  Rather, 
satellites should have formed in denser halos at high $z$.  The most luminous 
dSphs had $v_{max}$ up to 40km/s at infall (typically $z > 1$), and it is only 
through tidal stripping (aided by DM cored density profiles in the most luminous 
satellites) that they reach values below 25 km/s at $z=0$.  We predict that only 
satellite galaxies that have experienced substantial mass loss via tidal 
stripping should reach the lowest observed $v_c$ values at a fixed 
luminosity \citep[see also][]{Collins2013b}.

\section{Discussion}

\subsection{The Existence of Cores in dSphs}

The reduction in $v_c$ across all luminosities in our satellites simulated with 
baryonic physics is primarily due to tidal stripping, with the ability of baryons 
to cool to the central region of the host halo leading to stronger tidal forces 
on the satellites than in a DM-only run.  However, for the brightest satellites 
the mass loss is aided by the presence of a shallow DM density profile created 
by stellar feedback prior to accretion of the satellite onto the main halo.  
Below we discuss the observational evidence for cores in dSphs.

Most attempts to measure the central density slopes of dSphs have relied on using 
the spherical Jeans equations \citep[e.g.,][]{Gilmore2007}, but the mass and 
anisotropy of the stellar orbits are degenerate in the Jeans model, making the 
results highly dependent on adopted assumptions \citep{Evans2009}.  Recent works  
\citep{Wolf2012, Hayashi2012, Richardson2013} have attempted to overcome this issue 
by searching 
for maximum likelihoods in parameter space, but still adopt Jeans modeling.  
Schwarzchild modeling avoids some of the assumptions inherent in Jeans modeling, 
and has been applied to Fornax and Sculptor, with the conclusion that both have 
cored DM density profiles \citep{Jardel2012, Jardel2013, Jardel2013b, 
Breddels2013a}.  

Alternatively, dSphs with multiple stellar populations that span varying radii 
allow a direct fit to two independent derivations of the mass, avoiding assumptions 
about a mass model or isotropy.  Studies using this method \citep{Walker2011, 
Amorisco2012, Battaglia2008, Agnello2012} conclude that both Fornax (with slope 
measured out to $\sim$1 kpc) and Sculptor (measured interior to 500 pc) favor cored 
density profiles, and exclude cuspy density profiles at high significance.  It has 
also been argued that the fact that Fornax's globular cluster orbits have not 
decayed by dynamical friction requires a cored DM profile \citep{Goerdt2006, 
Cole2012}. A similar argument was recently made regarding observed stellar 
substructure in Sextans \citep{Lora2013}.

Overall, it has been argued both that the dSphs may all have a universal density 
profile \citep[e.g.,][]{Walker2009}, and that they don't \citep[e.g.,][]{Collins2013b}. 
There is also no accepted conclusion regarding whether any given dSph has 
a cuspy or cored density profile.  Despite the results above that favor a cored 
profile in Fornax, other studies have found it to be consistent with a cusp 
\citep{Strigari2010, Breddels2013b}.  This is not necessarily surprising given 
the difficulty and biases involved in modeling these systems \citep{Battaglia2013}.
For example, the results may depend sensitively on whether the line-of-sight views 
the minor or major axis of the dSph \citep{Kowalczyk2013, Laporte2013, Wojtak2013}.
Future galaxy-galaxy lensing experiments may be able to put additional 
constraints on the density profiles of satellites \citep{Li2013}.

The presence of cored density profiles at radii less than $\sim$ 500 pc in satellites 
fainter than $M_V \sim -12$ \citep[as seen in Sculptor by][]{Walker2011} is not 
ruled out by our simulations, as we do not resolve this region.  Although satellites 
in this luminosity range are unlikely to have had enough stellar feedback to produce 
large, kpc-sized cores \citep{Penarrubia2012, Garrison-kimmel2013}, it is possible 
that lower energies may lower the density profiles at much smaller radii. Higher 
resolution simulations are necessary to determine if small ($< 500$ pc) DM cores 
should exist in lower luminosity satellites. 

Despite the lack of conclusive evidence for cored profiles in dSphs, it is already 
clear that the overall concentrations of the MW dSph population are lower than 
predicted by CDM DM-only simulations \citep{Boylan-kolchin2012, Wolf2012, Hayashi2012}.  
The lowered densities of our satellites simulated with both DM and baryons are 
consistent with these observational results. 
The substantial reduction in the overall normalization of the central DM 
densities of satellites due to the presence of a disk will lead to a reduced DM 
annihilation signal.  Deeper searches for DM annihilation could help to constrain 
the central masses of these objects.

\subsection{Are Alternative DM Models Necessary?}

The discrepancies between various observational characteristics of the MW's dSph 
population and DM-only simulation predictions have led some theorists to suggest 
alternative models to CDM, such as Warm DM \citep[WDM,][]{Bode2001,Boyarsky2009, 
Tikhonov2009}
or self-interacting DM \citep[SIDM,][]{Spergel2000, Buckley2010, Tulin2013, 
Vogelsberger2013}.  We have shown in this paper that baryonic effects have the 
potential to resolve one of the problems of the CDM model, the ``too big to fail'' 
problem \citep{Boylan-kolchin2012}.  In \citet{Brooks2013} we argued that these 
same baryonic effects not only bring the masses, but also the numbers of massive 
satellites into agreement with observations, potentially solving the ``missing 
satellites'' problem \citep{Klypin1999,Moore1999}.  We discuss here the 
implications for a couple of popular alternative DM models.

First, we note that the effect of the enhanced tidal stripping in the 
presence of baryons may alter the mass and luminosity functions in a similar 
way to Warm DM (WDM) models. \citet{Nierenberg2013} examined both the satellite 
luminosity and mass functions of observed luminous satellites in galaxies with 
masses comparable to and greater than the MW between $0.1 < z < 0.8$.  They 
compared their observational results to the predicted mass and luminosity 
functions for three different CDM semi-analytic models, and one WDM semi-analytic 
model.  Despite varying physical prescriptions, none of the three CDM models was 
a good match to the data, which instead favored the WDM model results \citep[though 
see][]{Maccio2012b, Kang2013, Viel2013, Schneider2013}.  The WDM mass and luminosity 
functions were typically reduced by a factor $\sim$2-3 relative to the CDM predictions.  
This is consistent with our reductions when considering baryonic effects, though we 
caution that we examine only two halos.  Future work with a statistical sample of 
galaxies will put better constraints on the role of baryons.  However, the current 
results suggest that baryonic effects can reduce the number of subhalos in a comparable 
way to WDM models, requiring future work to identify predictions to rule out or 
favor a specific DM model.  In particular, we plan to run these same simulations 
in a WDM scenario with baryons, in order to quantify whether CDM+baryons can 
be observationally distinguished from WDM+baryons using dSphs. 

The primarily effect of WDM is to erase substructure below the free-streaming 
scale \citep{Angulo2013, Benson2013}.  Above this scale, the main difference of 
WDM halos from CDM halos is a lower concentration due to later formation times 
\citep{Destri2013,Lovell2013}. The later formation times yield SFHs that peak 
later in WDM models \citep{Herpich2013,Libeskind2013}.  It remains to be 
seen whether the presence of baryons yields a stronger effect in the lower 
concentrations of WDM halos.  However, SIDM should already lead to DM core 
formation, without the need for baryons.  Tension currently exists over the 
necessary cross-section for DM interaction that can yield core sizes as observed 
in dwarfs compared to halo shapes of galaxy clusters \citep{Peter2013, Rocha2013, 
Zavala2013}.  To date, no SIDM simulation has included baryons.  Again, 
future work is needed to test whether the combination of SIDM and baryons 
can reproduce observed trends in dSphs.

Before these simulations with WDM+baryons or SIDM+baryons can be completed, 
there are already hints that SIDM might be necessary in field dwarfs. As 
discussed above, dwarfs with stellar masses below $\sim$10$^7 M_{\odot}$ should 
not be able to create large DM density cores through stellar feedback.  Yet 
the observed $v_{rot}$ in galaxies at these low masses seems to be consistently 
lower than the expected $v_c$ in CDM DM-only simulations \citep{Papastergis2011, 
Ferrero2012}. If this discrepancy is not able to be reconciled by considering 
baryonic effects, or by a full consideration of observational biases, then it 
suggests some other mechanism might be necessary to reduce the central masses 
of these field dwarfs. SIDM would be a natural candidate.

\section{Conclusions}

We have demonstrated that simulations that account for the effects of SN feedback 
and enhanced tidal stripping on satellites result in a satellite population whose 
kinematic properties match the observed properties of the Milky Way and M31 satellites. 
Our findings are in sharp contrast to studies using DM-only  simulations, which 
over-predict the central masses of satellites in comparison to observations. 

By directly comparing the properties of simulated satellites that include gas 
hydrodynamics to the same satellites in DM-only simulations, we find:
\begin{enumerate}
\item The majority of satellites simulated with DM-only have $v_{max} > 20$ km/s, 
grossly inconsistent with the observed kinematics of the Local Group dSph population. 
We note that this is despite using simulated parent halos that are on the low side of 
the allowed range for the MW.  Although lowering the mass of the MW has been suggested 
as a solution to the ``too big to fail'' problem \citep{Vera-Ciro2013, Sawala2012}, 
our DM-only halos with virial masses of $7-8\times$10$^{11} M_{\odot}$ fail to produce 
a satellite population with kinematics consistent with observations.  In contrast, 
gas-free satellites simulated with baryonic physics have $v_{max}$ values in the range 
of $6-24$ km/s, matching the observed values of Local Group dSphs. The reduction in 
$v_{max}$ in these simulated galaxies is due to the combined effect of SN feedback on 
very luminous satellites (brighter than $M_V = -12$) and enhanced mass stripping for 
satellites across all luminosities in the presence of the parent galaxy's disk.
\item We find that the velocity dispersions of the simulated satellites are in good agreement with the range of observed dispersions in the MW and M31 dSph satellites. 
\item DM-only simulations produce satellites with $2-4$ times more mass in the central 
1 kpc than satellites simulated with baryonic physics. 
\item Satellites simulated with baryons and DM reproduce the observed scatter $v_c$ for 
dSphs, while satellites simulated with DM-only do not.  A tight $v_{max}$--luminosity 
relation exists for the satellites prior to infall.  After infall, the tidal effects 
of the baryonic disk in the host galaxy lead to large scatter based on the range of 
infall times and orbital pericenters of the satellites.  
\item Simulations that included baryonic physics have 1/3 fewer satellites that survive to z=0 than DM-only simulations. Six out of the eight DM-only satellites that have no surviving SPH counterpart have orbits that bring them within the central 30 kpc of their host galaxy. It seems therefore likely that the destruction of satellites in the baryonic run is due to tidal heating and shocking at the interface of the disk in the parent halo. These effects alter the predicted luminosity functions for satellites at $z=0$ from the DM-only case.   
\item Increased mass loss in tides for the satellites in the simulation with baryons, 
combined with total destruction of 1/3 of the satellites, shifts the velocity function 
of satellites at $z=0$ relative to the DM-only predictions.  The shift always acts to 
move satellites toward lower $v_{max}$ values than in the DM-only run.  At any given 
$v_{max}$, we find there are 50\% fewer satellites expected when baryonic effects are 
included.
\item Simulated field dwarf galaxies have systematically higher $v_{max}$ and 
$v_{\rm{1kpc}}$ values at $z=0$ than satellite galaxies in the same luminosity range.  
Tidal forces are necessary for satellites to reach the low $v_c$ values observed.
Preliminary results from observed field dwarfs suggest our simulated field dwarf 
velocities are consistent with observations in the local Universe. 
\item We find that simulated field dwarfs have similar velocity dispersions 
to simulated satellite galaxies when compared at the half light radii. These findings reproduce the recent observations of  \citep{Kirby2014}. 
\item Satellite galaxies have mass assembly histories that peak at higher redshifts
than isolated field dwarf galaxies in the same luminosity range.  These early 
assemblies are likely to be manifest in the cumulative SFHs of the dSphs vs dIrrs. 
\end{enumerate}

\acknowledgments
We thank Michael Boylan-Kolchin, Alan McConnachie, Louie Strigari, Matt Walker, 
and Beth Willman for useful discussions.  We thank Se-Heon Oh for supplying 
velocity data for Little THINGS galaxies. AB acknowledges support from The 
Grainger Foundation.  AZ acknowledges support from the Lady Davis Foundation.  
This material is based upon work supported in part by the National Science 
Foundation Grant No. 1066293 and the hospitality of the Aspen Center for 
Physics.  AZ's work was partially supported by the ISF grant 6/08, by GIF 
grant G-1052-104.7/2009, by the DFG  grant STE1869/1-1.GE625/15-1. 
Resources supporting this work were provided 
by the NASA High-End Computing (HEC) Program through the NASA Advanced 
Supercomputing (NAS) Division at Ames Research Center.  We thank Charlotte 
Christensen, Fabio Governato, Tom Quinn, Sijing Shen, and James Wadsley for 
use of the {\sc gasoline} code and simulations.


\begin{thebibliography}{141}
\expandafter\ifx\csname natexlab\endcsname\relax\def\natexlab#1{#1}\fi

\bibitem[{{Agertz} {et~al.}(2007){Agertz}, {Moore}, {Stadel}, {Potter},
  {Miniati}, {Read}, {Mayer}, {Gawryszczak}, {Kravtsov}, {Nordlund}, {Pearce},
  {Quilis}, {Rudd}, {Springel}, {Stone}, {Tasker}, {Teyssier}, {Wadsley}, \&
  {Walder}}]{Agertz2007}
{Agertz}, O. {et~al.} 2007, \mnras, 380, 963

\bibitem[{{Agnello} \& {Evans}(2012)}]{Agnello2012}
{Agnello}, A., \& {Evans}, N.~W. 2012, ArXiv e-prints

\bibitem[{{Amorisco} \& {Evans}(2012)}]{Amorisco2012}
{Amorisco}, N.~C., \& {Evans}, N.~W. 2012, \mnras, 419, 184

\bibitem[{{Anderhalden} {et~al.}(2013){Anderhalden}, {Schneider}, {Macci{\`o}},
  {Diemand}, \& {Bertone}}]{Anderhalden2013}
{Anderhalden}, D., {Schneider}, A., {Macci{\`o}}, A.~V., {Diemand}, J., \&
  {Bertone}, G. 2013, \jcap, 3, 14

\bibitem[{{Angulo} {et~al.}(2013){Angulo}, {Hahn}, \& {Abel}}]{Angulo2013}
{Angulo}, R.~E., {Hahn}, O., \& {Abel}, T. 2013, \mnras, 434, 3337

\bibitem[{{Arraki} {et~al.}(2012){Arraki}, {Klypin}, {More}, \&
  {Trujillo-Gomez}}]{Arraki2013}
{Arraki}, K.~S., {Klypin}, A., {More}, S., \& {Trujillo-Gomez}, S. 2012, ArXiv
  e-prints

\bibitem[{{Battaglia} {et~al.}(2013){Battaglia}, {Helmi}, \&
  {Breddels}}]{Battaglia2013}
{Battaglia}, G., {Helmi}, A., \& {Breddels}, M. 2013, ArXiv e-prints

\bibitem[{{Battaglia} {et~al.}(2008){Battaglia}, {Helmi}, {Tolstoy}, {Irwin},
  {Hill}, \& {Jablonka}}]{Battaglia2008}
{Battaglia}, G., {Helmi}, A., {Tolstoy}, E., {Irwin}, M., {Hill}, V., \&
  {Jablonka}, P. 2008, \apjl, 681, L13

\bibitem[{{Behroozi} {et~al.}(2010){Behroozi}, {Conroy}, \&
  {Wechsler}}]{Behroozi2010}
{Behroozi}, P.~S., {Conroy}, C., \& {Wechsler}, R.~H. 2010, \apj, 717, 379

\bibitem[{{Behroozi} {et~al.}(2013){Behroozi}, {Wechsler}, \&
  {Conroy}}]{Behroozi2013}
{Behroozi}, P.~S., {Wechsler}, R.~H., \& {Conroy}, C. 2013, \apj, 770, 57

\bibitem[{{Benson} {et~al.}(2013){Benson}, {Farahi}, {Cole}, {Moustakas},
  {Jenkins}, {Lovell}, {Kennedy}, {Helly}, \& {Frenk}}]{Benson2013}
{Benson}, A.~J. {et~al.} 2013, \mnras, 428, 1774

\bibitem[{{Bode} {et~al.}(2001){Bode}, {Ostriker}, \& {Turok}}]{Bode2001}
{Bode}, P., {Ostriker}, J.~P., \& {Turok}, N. 2001, \apj, 556, 93

\bibitem[{{Boyarsky} {et~al.}(2009){Boyarsky}, {Lesgourgues}, {Ruchayskiy}, \&
  {Viel}}]{Boyarsky2009}
{Boyarsky}, A., {Lesgourgues}, J., {Ruchayskiy}, O., \& {Viel}, M. 2009,
  Physical Review Letters, 102, 201304

\bibitem[{{Boylan-Kolchin} {et~al.}(2011){Boylan-Kolchin}, {Bullock}, \&
  {Kaplinghat}}]{Boylan-kolchin2011}
{Boylan-Kolchin}, M., {Bullock}, J.~S., \& {Kaplinghat}, M. 2011, \mnras, 415,
  L40

\bibitem[{{Boylan-Kolchin} {et~al.}(2012){Boylan-Kolchin}, {Bullock}, \&
  {Kaplinghat}}]{Boylan-kolchin2012}
---. 2012, \mnras, 2657

\bibitem[{{Breddels} \& {Helmi}(2013)}]{Breddels2013b}
{Breddels}, M.~A., \& {Helmi}, A. 2013, ArXiv e-prints

\bibitem[{{Breddels} {et~al.}(2013){Breddels}, {Helmi}, {van den Bosch}, {van
  de Ven}, \& {Battaglia}}]{Breddels2013a}
{Breddels}, M.~A., {Helmi}, A., {van den Bosch}, R.~C.~E., {van de Ven}, G., \&
  {Battaglia}, G. 2013, \mnras, 433, 3173

\bibitem[{{Brooks} {et~al.}(2013){Brooks}, {Kuhlen}, {Zolotov}, \&
  {Hooper}}]{Brooks2013}
{Brooks}, A.~M., {Kuhlen}, M., {Zolotov}, A., \& {Hooper}, D. 2013, \apj, 765,
  22

\bibitem[{{Buckley} \& {Fox}(2010)}]{Buckley2010}
{Buckley}, M.~R., \& {Fox}, P.~J. 2010, \prd, 81, 083522

\bibitem[{{Chang} {et~al.}(2013){Chang}, {Macci{\`o}}, \& {Kang}}]{Chang2013}
{Chang}, J., {Macci{\`o}}, A.~V., \& {Kang}, X. 2013, \mnras, 431, 3533

\bibitem[{{Choi} {et~al.}(2009){Choi}, {Weinberg}, \& {Katz}}]{Choi2009}
{Choi}, J.-H., {Weinberg}, M.~D., \& {Katz}, N. 2009, \mnras, 400, 1247

\bibitem[{{Christensen} {et~al.}(2012){Christensen}, {Quinn}, {Governato},
  {Stilp}, {Shen}, \& {Wadsley}}]{Christensen2012}
{Christensen}, C., {Quinn}, T., {Governato}, F., {Stilp}, A., {Shen}, S., \&
  {Wadsley}, J. 2012, ArXiv e-prints

\bibitem[{{Cole} {et~al.}(2012){Cole}, {Dehnen}, {Read}, \&
  {Wilkinson}}]{Cole2012}
{Cole}, D.~R., {Dehnen}, W., {Read}, J.~I., \& {Wilkinson}, M.~I. 2012, ArXiv
  e-prints

\bibitem[{{Collins} {et~al.}(2013{\natexlab{a}}){Collins}, {Chapman}, {Rich},
  {Ibata}, {Martin}, {Irwin}, {Bate}, {Lewis}, {Pe{\~n}arrubia}, {Arimoto},
  {Casey}, {Ferguson}, {Koch}, {McConnachie}, \& {Tanvir}}]{Collins2013}
{Collins}, M.~L.~M. {et~al.} 2013{\natexlab{a}}, \apj, 768, 172

\bibitem[{{Collins} {et~al.}(2013{\natexlab{b}}){Collins}, {Chapman}, {Rich},
  {Ibata}, {Martin}, {Irwin}, {Bate}, {Lewis}, {Pe{\~n}arrubia}, {Arimoto},
  {Casey}, {Ferguson}, {Koch}, {McConnachie}, \& {Tanvir}}]{Collins2013b}
---. 2013{\natexlab{b}}, ArXiv e-prints

\bibitem[{{Collins} {et~al.}(2011){Collins}, {Chapman}, {Rich}, {Irwin},
  {Pe{\~n}arrubia}, {Ibata}, {Arimoto}, {Brooks}, {Ferguson}, {Lewis},
  {McConnachie}, \& {Venn}}]{Collins2011}
---. 2011, \mnras, 417, 1170

\bibitem[{{C{\^o}t{\'e}} {et~al.}(2000){C{\^o}t{\'e}}, {Carignan}, \&
  {Freeman}}]{Cote2000}
{C{\^o}t{\'e}}, S., {Carignan}, C., \& {Freeman}, K.~C. 2000, \aj, 120, 3027

\bibitem[{{de Boer} {et~al.}(2013){de Boer}, {Tolstoy}, {Saha}, \&
  {Olszewski}}]{deBoer2013}
{de Boer}, T.~J.~L., {Tolstoy}, E., {Saha}, A., \& {Olszewski}, E.~W. 2013,
  \aap, 551, A103

\bibitem[{{Del Popolo}(2009)}]{Delpopolo2009}
{Del Popolo}, A. 2009, \apj, 698, 2093

\bibitem[{{Dellenbusch} {et~al.}(2008){Dellenbusch}, {Gallagher}, {Knezek}, \&
  {Noble}}]{Dellenbusch2008}
{Dellenbusch}, K.~E., {Gallagher}, III, J.~S., {Knezek}, P.~M., \& {Noble},
  A.~G. 2008, \aj, 135, 326

\bibitem[{{Destri} {et~al.}(2013){Destri}, {de Vega}, \&
  {Sanchez}}]{Destri2013}
{Destri}, C., {de Vega}, H.~J., \& {Sanchez}, N.~G. 2013, ArXiv e-prints

\bibitem[{{Diemand} {et~al.}(2007){Diemand}, {Kuhlen}, \&
  {Madau}}]{Diemand2007}
{Diemand}, J., {Kuhlen}, M., \& {Madau}, P. 2007, \apj, 667, 859

\bibitem[{{Diemand} {et~al.}(2005){Diemand}, {Madau}, \& {Moore}}]{Diemand2005}
{Diemand}, J., {Madau}, P., \& {Moore}, B. 2005, \mnras, 364, 367

\bibitem[{{D'Onghia} {et~al.}(2010){D'Onghia}, {Springel}, {Hernquist}, \&
  {Keres}}]{D'Onghia2010}
{D'Onghia}, E., {Springel}, V., {Hernquist}, L., \& {Keres}, D. 2010, \apj,
  709, 1138

\bibitem[{{El-Zant} {et~al.}(2001){El-Zant}, {Shlosman}, \&
  {Hoffman}}]{ElZant2001}
{El-Zant}, A., {Shlosman}, I., \& {Hoffman}, Y. 2001, \apj, 560, 636

\bibitem[{{Evans} {et~al.}(2009){Evans}, {An}, \& {Walker}}]{Evans2009}
{Evans}, N.~W., {An}, J., \& {Walker}, M.~G. 2009, \mnras, 393, L50

\bibitem[{{Ferrero} {et~al.}(2012){Ferrero}, {Abadi}, {Navarro}, {Sales}, \&
  {Gurovich}}]{Ferrero2012}
{Ferrero}, I., {Abadi}, M.~G., {Navarro}, J.~F., {Sales}, L.~V., \& {Gurovich},
  S. 2012, \mnras, 425, 2817

\bibitem[{{Garrison-Kimmel} {et~al.}(2013){Garrison-Kimmel}, {Rocha},
  {Boylan-Kolchin}, {Bullock}, \& {Lally}}]{Garrison-kimmel2013}
{Garrison-Kimmel}, S., {Rocha}, M., {Boylan-Kolchin}, M., {Bullock}, J.~S., \&
  {Lally}, J. 2013, \mnras

\bibitem[{{Gill} {et~al.}(2004){Gill}, {Knebe}, \& {Gibson}}]{Gill2004}
{Gill}, S.~P.~D., {Knebe}, A., \& {Gibson}, B.~K. 2004, \mnras, 351, 399

\bibitem[{{Gilmore} {et~al.}(2007){Gilmore}, {Wilkinson}, {Kleyna}, {Koch},
  {Evans}, {Wyse}, \& {Grebel}}]{Gilmore2007}
{Gilmore}, G., {Wilkinson}, M., {Kleyna}, J., {Koch}, A., {Evans}, W., {Wyse},
  R.~F.~G., \& {Grebel}, E.~K. 2007, Nuclear Physics B Proceedings Supplements,
  173, 15

\bibitem[{{Gnedin} {et~al.}(1999){Gnedin}, {Hernquist}, \&
  {Ostriker}}]{Gnedin1999}
{Gnedin}, O.~Y., {Hernquist}, L., \& {Ostriker}, J.~P. 1999, \apj, 514, 109

\bibitem[{{Goerdt} {et~al.}(2006){Goerdt}, {Moore}, {Read}, {Stadel}, \&
  {Zemp}}]{Goerdt2006}
{Goerdt}, T., {Moore}, B., {Read}, J.~I., {Stadel}, J., \& {Zemp}, M. 2006,
  \mnras, 368, 1073

\bibitem[{{Governato} {et~al.}(2012){Governato}, {Zolotov}, {Pontzen},
  {Christensen}, {Oh}, {Brooks}, {Quinn}, {Shen}, \& {Wadsley}}]{Governato2012}
{Governato}, F. {et~al.} 2012, ArXiv e-prints

\bibitem[{{Grebel} \& {Gallagher}(2004)}]{Grebel2004}
{Grebel}, E.~K., \& {Gallagher}, III, J.~S. 2004, \apjl, 610, L89

\bibitem[{{Haardt} \& {Madau}(2001)}]{Haardt2001}
{Haardt}, F., \& {Madau}, P. 2001, in Clusters of Galaxies and the High
  Redshift Universe Observed in X-rays, ed. {D.~M.~Neumann \& J.~T.~V.~Tran}

\bibitem[{{Hayashi} {et~al.}(2003){Hayashi}, {Navarro}, {Taylor}, {Stadel}, \&
  {Quinn}}]{Hayashi2003}
{Hayashi}, E., {Navarro}, J.~F., {Taylor}, J.~E., {Stadel}, J., \& {Quinn}, T.
  2003, \apj, 584, 541

\bibitem[{{Hayashi} \& {Chiba}(2012)}]{Hayashi2012}
{Hayashi}, K., \& {Chiba}, M. 2012, \apj, 755, 145

\bibitem[{{Hearin} {et~al.}(2013){Hearin}, {Zentner}, {Berlind}, \&
  {Newman}}]{Hearin2013}
{Hearin}, A.~P., {Zentner}, A.~R., {Berlind}, A.~A., \& {Newman}, J.~A. 2013,
  \mnras, 433, 659

\bibitem[{{Herpich} {et~al.}(2013){Herpich}, {Stinson}, {Macci{\`o}}, {Brook},
  {Wadsley}, {Couchman}, \& {Quinn}}]{Herpich2013}
{Herpich}, J., {Stinson}, G.~S., {Macci{\`o}}, A.~V., {Brook}, C., {Wadsley},
  J., {Couchman}, H.~M.~P., \& {Quinn}, T. 2013, ArXiv e-prints

\bibitem[{{Jardel} \& {Gebhardt}(2012)}]{Jardel2012}
{Jardel}, J.~R., \& {Gebhardt}, K. 2012, \apj, 746, 89

\bibitem[{{Jardel} \& {Gebhardt}(2013)}]{Jardel2013b}
---. 2013, ArXiv e-prints

\bibitem[{{Jardel} {et~al.}(2013){Jardel}, {Gebhardt}, {Fabricius}, {Drory}, \&
  {Williams}}]{Jardel2013}
{Jardel}, J.~R., {Gebhardt}, K., {Fabricius}, M.~H., {Drory}, N., \&
  {Williams}, M.~J. 2013, \apj, 763, 91

\bibitem[{{Kang} {et~al.}(2013){Kang}, {Macci{\`o}}, \& {Dutton}}]{Kang2013}
{Kang}, X., {Macci{\`o}}, A.~V., \& {Dutton}, A.~A. 2013, \apj, 767, 22

\bibitem[{{Kazantzidis} {et~al.}(2004){Kazantzidis}, {Mayer}, {Mastropietro},
  {Diemand}, {Stadel}, \& {Moore}}]{Kazantzidis2004}
{Kazantzidis}, S., {Mayer}, L., {Mastropietro}, C., {Diemand}, J., {Stadel},
  J., \& {Moore}, B. 2004, \apj, 608, 663

\bibitem[{{Kirby} {et~al.}(2014){Kirby}, {Bullock}, {Boylan-Kolchin},
  {Kaplinghat}, \& {Cohen}}]{Kirby2014}
{Kirby}, E.~N., {Bullock}, J.~S., {Boylan-Kolchin}, M., {Kaplinghat}, M., \&
  {Cohen}, J.~G. 2014, \mnras

\bibitem[{{Klypin} {et~al.}(1999){Klypin}, {Kravtsov}, {Valenzuela}, \&
  {Prada}}]{Klypin1999}
{Klypin}, A., {Kravtsov}, A.~V., {Valenzuela}, O., \& {Prada}, F. 1999, \apj,
  522, 82

\bibitem[{{Knollmann} \& {Knebe}(2009)}]{Knollmann2009}
{Knollmann}, S.~R., \& {Knebe}, A. 2009, \apjs, 182, 608

\bibitem[{{Koposov} {et~al.}(2009){Koposov}, {Yoo}, {Rix}, {Weinberg},
  {Macci{\`o}}, \& {Escud{\'e}}}]{Koposov2009}
{Koposov}, S.~E., {Yoo}, J., {Rix}, H.-W., {Weinberg}, D.~H., {Macci{\`o}},
  A.~V., \& {Escud{\'e}}, J.~M. 2009, \apj, 696, 2179

\bibitem[{{Kowalczyk} {et~al.}(2013){Kowalczyk}, {{\L}okas}, {Kazantzidis}, \&
  {Mayer}}]{Kowalczyk2013}
{Kowalczyk}, K., {{\L}okas}, E.~L., {Kazantzidis}, S., \& {Mayer}, L. 2013,
  \mnras, 431, 2796

\bibitem[{{Kravtsov}(2010)}]{Kravtsov2010}
{Kravtsov}, A. 2010, Advances in Astronomy, 2010

\bibitem[{{Kravtsov} {et~al.}(2004){Kravtsov}, {Gnedin}, \&
  {Klypin}}]{Kravtsov2004}
{Kravtsov}, A.~V., {Gnedin}, O.~Y., \& {Klypin}, A.~A. 2004, \apj, 609, 482

\bibitem[{{Laporte} {et~al.}(2013){Laporte}, {Walker}, \&
  {Pe{\~n}arrubia}}]{Laporte2013}
{Laporte}, C.~F.~P., {Walker}, M.~G., \& {Pe{\~n}arrubia}, J. 2013, \mnras,
  433, L54

\bibitem[{{Leauthaud} {et~al.}(2012){Leauthaud}, {George}, {Behroozi}, {Bundy},
  {Tinker}, {Wechsler}, {Conroy}, {Finoguenov}, \& {Tanaka}}]{Leauthaud2012}
{Leauthaud}, A. {et~al.} 2012, \apj, 746, 95

\bibitem[{{Li} {et~al.}(2013){Li}, {Mo}, {Fan}, {Yang}, \& {Bosch}}]{Li2013}
{Li}, R., {Mo}, H.~J., {Fan}, Z., {Yang}, X., \& {Bosch}, F.~C.~v.~d. 2013,
  \mnras, 430, 3359

\bibitem[{{Libeskind} {et~al.}(2013){Libeskind}, {Di Cintio}, {Knebe}, {Yepes},
  {Gottl{\"o}ber}, {Steinmetz}, {Hoffman}, \&
  {Martinez-Vaquero}}]{Libeskind2013}
{Libeskind}, N.~I., {Di Cintio}, A., {Knebe}, A., {Yepes}, G., {Gottl{\"o}ber},
  S., {Steinmetz}, M., {Hoffman}, Y., \& {Martinez-Vaquero}, L.~A. 2013, \pasa,
  30, 39

\bibitem[{{Libeskind} {et~al.}(2011){Libeskind}, {Knebe}, {Hoffman},
  {Gottl{\"o}ber}, \& {Yepes}}]{Libeskind2011}
{Libeskind}, N.~I., {Knebe}, A., {Hoffman}, Y., {Gottl{\"o}ber}, S., \&
  {Yepes}, G. 2011, \mnras, 418, 336

\bibitem[{{Lora} {et~al.}(2013){Lora}, {Grebel}, {Sanchez-Salcedo}, \&
  {Just}}]{Lora2013}
{Lora}, V., {Grebel}, E.~K., {Sanchez-Salcedo}, F.~J., \& {Just}, A. 2013,
  ArXiv e-prints

\bibitem[{{Lovell} {et~al.}(2012){Lovell}, {Eke}, {Frenk}, {Gao}, {Jenkins},
  {Theuns}, {Wang}, {White}, {Boyarsky}, \& {Ruchayskiy}}]{Lovell2012}
{Lovell}, M.~R. {et~al.} 2012, \mnras, 420, 2318

\bibitem[{{Lovell} {et~al.}(2013){Lovell}, {Frenk}, {Eke}, {Jenkins}, {Gao}, \&
  {Theuns}}]{Lovell2013}
{Lovell}, M.~R., {Frenk}, C.~S., {Eke}, V.~R., {Jenkins}, A., {Gao}, L., \&
  {Theuns}, T. 2013, ArXiv e-prints

\bibitem[{{Macci{\`o}} \& {Fontanot}(2010)}]{Maccio2010}
{Macci{\`o}}, A.~V., \& {Fontanot}, F. 2010, \mnras, 404, L16

\bibitem[{{Macci{\`o}} {et~al.}(2012{\natexlab{a}}){Macci{\`o}}, {Paduroiu},
  {Anderhalden}, {Schneider}, \& {Moore}}]{Maccio2012b}
{Macci{\`o}}, A.~V., {Paduroiu}, S., {Anderhalden}, D., {Schneider}, A., \&
  {Moore}, B. 2012{\natexlab{a}}, \mnras, 424, 1105

\bibitem[{{Macci{\`o}} {et~al.}(2012{\natexlab{b}}){Macci{\`o}}, {Stinson},
  {Brook}, {Wadsley}, {Couchman}, {Shen}, {Gibson}, \& {Quinn}}]{Maccio2012}
{Macci{\`o}}, A.~V., {Stinson}, G., {Brook}, C.~B., {Wadsley}, J., {Couchman},
  H.~M.~P., {Shen}, S., {Gibson}, B.~K., \& {Quinn}, T. 2012{\natexlab{b}},
  \apjl, 744, L9

\bibitem[{{Madau} {et~al.}(2008){Madau}, {Diemand}, \& {Kuhlen}}]{Madau2008}
{Madau}, P., {Diemand}, J., \& {Kuhlen}, M. 2008, \apj, 679, 1260

\bibitem[{{Martinez}(2013)}]{Martinez2013}
{Martinez}, G.~D. 2013, ArXiv e-prints

\bibitem[{{Mateo}(1998)}]{Mateo1998}
{Mateo}, M.~L. 1998, \araa, 36, 435

\bibitem[{{McConnachie}(2012)}]{McConnachie2012}
{McConnachie}, A.~W. 2012, \aj, 144, 4

\bibitem[{{McGaugh} {et~al.}(2007){McGaugh}, {de Blok}, {Schombert}, {Kuzio de
  Naray}, \& {Kim}}]{McGaugh2007}
{McGaugh}, S.~S., {de Blok}, W.~J.~G., {Schombert}, J.~M., {Kuzio de Naray},
  R., \& {Kim}, J.~H. 2007, \apj, 659, 149

\bibitem[{{Moore} {et~al.}(2006){Moore}, {Diemand}, {Madau}, {Zemp}, \&
  {Stadel}}]{Moore2006}
{Moore}, B., {Diemand}, J., {Madau}, P., {Zemp}, M., \& {Stadel}, J. 2006,
  \mnras, 368, 563

\bibitem[{{Moore} {et~al.}(1999){Moore}, {Ghigna}, {Governato}, {Lake},
  {Quinn}, {Stadel}, \& {Tozzi}}]{Moore1999}
{Moore}, B., {Ghigna}, S., {Governato}, F., {Lake}, G., {Quinn}, T., {Stadel},
  J., \& {Tozzi}, P. 1999, \apjl, 524, L19

\bibitem[{{Moster} {et~al.}(2012){Moster}, {Naab}, \& {White}}]{Moster2012}
{Moster}, B.~P., {Naab}, T., \& {White}, S.~D.~M. 2012, ArXiv e-prints

\bibitem[{{Mu{\~n}oz} {et~al.}(2008){Mu{\~n}oz}, {Majewski}, \&
  {Johnston}}]{Munoz2008}
{Mu{\~n}oz}, R.~R., {Majewski}, S.~R., \& {Johnston}, K.~V. 2008, \apj, 679,
  346

\bibitem[{{Nichols} \& {Bland-Hawthorn}(2013)}]{Nichols2013}
{Nichols}, M., \& {Bland-Hawthorn}, J. 2013, \apj, 775, 97

\bibitem[{{Nierenberg} {et~al.}(2013){Nierenberg}, {Treu}, {Menci}, {Lu}, \&
  {Wang}}]{Nierenberg2013}
{Nierenberg}, A.~M., {Treu}, T., {Menci}, N., {Lu}, Y., \& {Wang}, W. 2013,
  \apj, 772, 146

\bibitem[{{Oh} {et~al.}(2011){Oh}, {Brook}, {Governato}, {Brinks}, {Mayer}, {de
  Blok}, {Brooks}, \& {Walter}}]{Oh2011}
{Oh}, S.-H., {Brook}, C., {Governato}, F., {Brinks}, E., {Mayer}, L., {de
  Blok}, W.~J.~G., {Brooks}, A., \& {Walter}, F. 2011, \aj, 142, 24

\bibitem[{{Okamoto} {et~al.}(2008){Okamoto}, {Gao}, \& {Theuns}}]{Okamoto2008}
{Okamoto}, T., {Gao}, L., \& {Theuns}, T. 2008, \mnras, 390, 920

\bibitem[{{Oman} {et~al.}(2013){Oman}, {Hudson}, \& {Behroozi}}]{Oman2013}
{Oman}, K.~A., {Hudson}, M.~J., \& {Behroozi}, P.~S. 2013, \mnras, 431, 2307

\bibitem[{{Ostriker} {et~al.}(1972){Ostriker}, {Spitzer}, \&
  {Chevalier}}]{Ostriker1972}
{Ostriker}, J.~P., {Spitzer}, Jr., L., \& {Chevalier}, R.~A. 1972, \apjl, 176,
  L51

\bibitem[{{Papastergis} {et~al.}(2011){Papastergis}, {Martin}, {Giovanelli}, \&
  {Haynes}}]{Papastergis2011}
{Papastergis}, E., {Martin}, A.~M., {Giovanelli}, R., \& {Haynes}, M.~P. 2011,
  \apj, 739, 38

\bibitem[{{Pe{\~n}arrubia} {et~al.}(2010){Pe{\~n}arrubia}, {Benson}, {Walker},
  {Gilmore}, {McConnachie}, \& {Mayer}}]{Penarrubia2010}
{Pe{\~n}arrubia}, J., {Benson}, A.~J., {Walker}, M.~G., {Gilmore}, G.,
  {McConnachie}, A.~W., \& {Mayer}, L. 2010, \mnras, 406, 1290

\bibitem[{{Pe{\~n}arrubia} {et~al.}(2008){Pe{\~n}arrubia}, {McConnachie}, \&
  {Navarro}}]{Penarrubia2008}
{Pe{\~n}arrubia}, J., {McConnachie}, A.~W., \& {Navarro}, J.~F. 2008, \apj,
  672, 904

\bibitem[{{Pe{\~n}arrubia} {et~al.}(2012){Pe{\~n}arrubia}, {Pontzen}, {Walker},
  \& {Koposov}}]{Penarrubia2012}
{Pe{\~n}arrubia}, J., {Pontzen}, A., {Walker}, M.~G., \& {Koposov}, S.~E. 2012,
  \apjl, 759, L42

\bibitem[{{Peter} {et~al.}(2013){Peter}, {Rocha}, {Bullock}, \&
  {Kaplinghat}}]{Peter2013}
{Peter}, A.~H.~G., {Rocha}, M., {Bullock}, J.~S., \& {Kaplinghat}, M. 2013,
  \mnras, 430, 105

\bibitem[{{Polisensky} \& {Ricotti}(2013)}]{Polisensky2013}
{Polisensky}, E., \& {Ricotti}, M. 2013, ArXiv e-prints

\bibitem[{{Pontzen} \& {Governato}(2012)}]{Pontzen2012}
{Pontzen}, A., \& {Governato}, F. 2012, \mnras, 421, 3464

\bibitem[{{Rashkov} {et~al.}(2012){Rashkov}, {Madau}, {Kuhlen}, \&
  {Diemand}}]{Rashkov2012}
{Rashkov}, V., {Madau}, P., {Kuhlen}, M., \& {Diemand}, J. 2012, \apj, 745, 142

\bibitem[{{Read} \& {Gilmore}(2005)}]{Read2005}
{Read}, J.~I., \& {Gilmore}, G. 2005, \mnras, 356, 107

\bibitem[{{Read} {et~al.}(2006){Read}, {Wilkinson}, {Evans}, {Gilmore}, \&
  {Kleyna}}]{Read2006b}
{Read}, J.~I., {Wilkinson}, M.~I., {Evans}, N.~W., {Gilmore}, G., \& {Kleyna},
  J.~T. 2006, \mnras, 367, 387

\bibitem[{{Richardson} \& {Fairbairn}(2013)}]{Richardson2013}
{Richardson}, T., \& {Fairbairn}, M. 2013, ArXiv e-prints

\bibitem[{{Rocha} {et~al.}(2013){Rocha}, {Peter}, {Bullock}, {Kaplinghat},
  {Garrison-Kimmel}, {O{\~n}orbe}, \& {Moustakas}}]{Rocha2013}
{Rocha}, M., {Peter}, A.~H.~G., {Bullock}, J.~S., {Kaplinghat}, M.,
  {Garrison-Kimmel}, S., {O{\~n}orbe}, J., \& {Moustakas}, L.~A. 2013, \mnras,
  430, 81

\bibitem[{{Rodr{\'{\i}}guez-Puebla} {et~al.}(2013){Rodr{\'{\i}}guez-Puebla},
  {Avila-Reese}, \& {Drory}}]{Rodriguez-Puebla2013}
{Rodr{\'{\i}}guez-Puebla}, A., {Avila-Reese}, V., \& {Drory}, N. 2013, \apj,
  773, 172

\bibitem[{{Romano-D{\'{\i}}az} {et~al.}(2010){Romano-D{\'{\i}}az}, {Shlosman},
  {Heller}, \& {Hoffman}}]{Romano2010}
{Romano-D{\'{\i}}az}, E., {Shlosman}, I., {Heller}, C., \& {Hoffman}, Y. 2010,
  \apj, 716, 1095

\bibitem[{{Sales} {et~al.}(2007{\natexlab{a}}){Sales}, {Navarro}, {Abadi}, \&
  {Steinmetz}}]{Sales2007b}
{Sales}, L.~V., {Navarro}, J.~F., {Abadi}, M.~G., \& {Steinmetz}, M.
  2007{\natexlab{a}}, \mnras, 379, 1475

\bibitem[{{Sales} {et~al.}(2007{\natexlab{b}}){Sales}, {Navarro}, {Abadi}, \&
  {Steinmetz}}]{Sales2007}
---. 2007{\natexlab{b}}, \mnras, 379, 1464

\bibitem[{{Sales} {et~al.}(2013){Sales}, {Wang}, {White}, \&
  {Navarro}}]{Sales2013}
{Sales}, L.~V., {Wang}, W., {White}, S.~D.~M., \& {Navarro}, J.~F. 2013,
  \mnras, 428, 573

\bibitem[{{Sawala} {et~al.}(2012){Sawala}, {Scannapieco}, \&
  {White}}]{Sawala2012}
{Sawala}, T., {Scannapieco}, C., \& {White}, S. 2012, \mnras, 420, 1714

\bibitem[{{Schewtschenko} \& {Macci{\`o}}(2011)}]{Schewtschenko2011}
{Schewtschenko}, J.~A., \& {Macci{\`o}}, A.~V. 2011, \mnras, 413, 878

\bibitem[{{Schneider} {et~al.}(2013){Schneider}, {Anderhalden}, {Maccio}, \&
  {Diemand}}]{Schneider2013}
{Schneider}, A., {Anderhalden}, D., {Maccio}, A., \& {Diemand}, J. 2013, ArXiv
  e-prints

\bibitem[{{Shao} {et~al.}(2013){Shao}, {Gao}, {Theuns}, \& {Frenk}}]{Shao2013}
{Shao}, S., {Gao}, L., {Theuns}, T., \& {Frenk}, C.~S. 2013, \mnras, 430, 2346

\bibitem[{{Shen} {et~al.}(2010){Shen}, {Wadsley}, \& {Stinson}}]{Shen2010}
{Shen}, S., {Wadsley}, J., \& {Stinson}, G. 2010, \mnras, 1043

\bibitem[{{Simon} \& {Geha}(2007)}]{Simon2007}
{Simon}, J.~D., \& {Geha}, M. 2007, \apj, 670, 313

\bibitem[{{Spergel} \& {Steinhardt}(2000)}]{Spergel2000}
{Spergel}, D.~N., \& {Steinhardt}, P.~J. 2000, Physical Review Letters, 84,
  3760

\bibitem[{{Springel} {et~al.}(2008){Springel}, {Wang}, {Vogelsberger},
  {Ludlow}, {Jenkins}, {Helmi}, {Navarro}, {Frenk}, \& {White}}]{Springel2008}
{Springel}, V. {et~al.} 2008, \mnras, 391, 1685

\bibitem[{{Stadel}(2001)}]{Stadel2001}
{Stadel}, J.~G. 2001, PhD thesis, UNIVERSITY OF WASHINGTON

\bibitem[{{Stinson} {et~al.}(2006){Stinson}, {Seth}, {Katz}, {Wadsley},
  {Governato}, \& {Quinn}}]{Stinson2006}
{Stinson}, G., {Seth}, A., {Katz}, N., {Wadsley}, J., {Governato}, F., \&
  {Quinn}, T. 2006, \mnras, 373, 1074

\bibitem[{{Strigari} {et~al.}(2010){Strigari}, {Frenk}, \&
  {White}}]{Strigari2010}
{Strigari}, L.~E., {Frenk}, C.~S., \& {White}, S.~D.~M. 2010, \mnras, 408, 2364

\bibitem[{{Taylor} \& {Babul}(2001)}]{Taylor2001}
{Taylor}, J.~E., \& {Babul}, A. 2001, \apj, 559, 716

\bibitem[{{Teyssier} {et~al.}(2012{\natexlab{a}}){Teyssier}, {Johnston}, \&
  {Kuhlen}}]{MTeyssier2012}
{Teyssier}, M., {Johnston}, K.~V., \& {Kuhlen}, M. 2012{\natexlab{a}}, \mnras,
  426, 1808

\bibitem[{{Teyssier} {et~al.}(2012{\natexlab{b}}){Teyssier}, {Pontzen},
  {Dubois}, \& {Read}}]{Teyssier2012}
{Teyssier}, R., {Pontzen}, A., {Dubois}, Y., \& {Read}, J. 2012{\natexlab{b}},
  ArXiv e-prints

\bibitem[{{Tikhonov} {et~al.}(2009){Tikhonov}, {Gottl{\"o}ber}, {Yepes}, \&
  {Hoffman}}]{Tikhonov2009}
{Tikhonov}, A.~V., {Gottl{\"o}ber}, S., {Yepes}, G., \& {Hoffman}, Y. 2009,
  \mnras, 399, 1611

\bibitem[{{Tollerud} {et~al.}(2012){Tollerud}, {Beaton}, {Geha}, {Bullock},
  {Guhathakurta}, {Kalirai}, {Majewski}, {Kirby}, {Gilbert}, {Yniguez},
  {Patterson}, {Ostheimer}, {Cooke}, {Dorman}, {Choudhury}, \&
  {Cooper}}]{Tollerud2012}
{Tollerud}, E.~J. {et~al.} 2012, \apj, 752, 45

\bibitem[{{Tollerud} {et~al.}(2008){Tollerud}, {Bullock}, {Strigari}, \&
  {Willman}}]{Tollerud2008}
{Tollerud}, E.~J., {Bullock}, J.~S., {Strigari}, L.~E., \& {Willman}, B. 2008,
  \apj, 688, 277

\bibitem[{{Tollerud} {et~al.}(2013){Tollerud}, {Geha}, {Vargas}, \&
  {Bullock}}]{Tollerud2013}
{Tollerud}, E.~J., {Geha}, M.~C., {Vargas}, L.~C., \& {Bullock}, J.~S. 2013,
  \apj, 768, 50

\bibitem[{{Tonini} {et~al.}(2006){Tonini}, {Lapi}, \& {Salucci}}]{Tonini2006}
{Tonini}, C., {Lapi}, A., \& {Salucci}, P. 2006, \apj, 649, 591

\bibitem[{{Tulin} {et~al.}(2013){Tulin}, {Yu}, \& {Zurek}}]{Tulin2013}
{Tulin}, S., {Yu}, H.-B., \& {Zurek}, K.~M. 2013, \prd, 87, 115007

\bibitem[{{Vera-Ciro} {et~al.}(2013){Vera-Ciro}, {Helmi}, {Starkenburg}, \&
  {Breddels}}]{Vera-Ciro2013}
{Vera-Ciro}, C.~A., {Helmi}, A., {Starkenburg}, E., \& {Breddels}, M.~A. 2013,
  \mnras, 428, 1696

\bibitem[{{Viel} {et~al.}(2013){Viel}, {Becker}, {Bolton}, \&
  {Haehnelt}}]{Viel2013}
{Viel}, M., {Becker}, G.~D., {Bolton}, J.~S., \& {Haehnelt}, M.~G. 2013, \prd,
  88, 043502

\bibitem[{{Vogelsberger} \& {Zavala}(2013)}]{Vogelsberger2013}
{Vogelsberger}, M., \& {Zavala}, J. 2013, \mnras, 430, 1722

\bibitem[{{Vogelsberger} {et~al.}(2012){Vogelsberger}, {Zavala}, \&
  {Loeb}}]{Vogelsberger2012}
{Vogelsberger}, M., {Zavala}, J., \& {Loeb}, A. 2012, \mnras, 423, 3740

\bibitem[{{Wadepuhl} \& {Springel}(2011)}]{Wadepuhl2011}
{Wadepuhl}, M., \& {Springel}, V. 2011, \mnras, 410, 1975

\bibitem[{{Wadsley} {et~al.}(2004){Wadsley}, {Stadel}, \&
  {Quinn}}]{Wadsley2004}
{Wadsley}, J.~W., {Stadel}, J., \& {Quinn}, T. 2004, New Astronomy, 9, 137

\bibitem[{{Walker} {et~al.}(2009){Walker}, {Mateo}, {Olszewski},
  {Pe{\~n}arrubia}, {Wyn Evans}, \& {Gilmore}}]{Walker2009}
{Walker}, M.~G., {Mateo}, M., {Olszewski}, E.~W., {Pe{\~n}arrubia}, J., {Wyn
  Evans}, N., \& {Gilmore}, G. 2009, \apj, 704, 1274

\bibitem[{{Walker} \& {Pe{\~n}arrubia}(2011)}]{Walker2011}
{Walker}, M.~G., \& {Pe{\~n}arrubia}, J. 2011, \apj, 742, 20

\bibitem[{{Walsh} {et~al.}(2009){Walsh}, {Willman}, \& {Jerjen}}]{Walsh2009}
{Walsh}, S.~M., {Willman}, B., \& {Jerjen}, H. 2009, \aj, 137, 450

\bibitem[{{Wechsler} {et~al.}(2002){Wechsler}, {Bullock}, {Primack},
  {Kravtsov}, \& {Dekel}}]{Wechsler2002}
{Wechsler}, R.~H., {Bullock}, J.~S., {Primack}, J.~R., {Kravtsov}, A.~V., \&
  {Dekel}, A. 2002, \apj, 568, 52

\bibitem[{{Weisz} {et~al.}(2011){Weisz}, {Dalcanton}, {Williams}, {Gilbert},
  {Skillman}, {Seth}, {Dolphin}, {McQuinn}, {Gogarten}, {Holtzman}, {Rosema},
  {Cole}, {Karachentsev}, \& {Zaritsky}}]{Weisz2011}
{Weisz}, D.~R. {et~al.} 2011, \apj, 739, 5

\bibitem[{{Wojtak} {et~al.}(2013){Wojtak}, {Gottl{\"o}ber}, \&
  {Klypin}}]{Wojtak2013}
{Wojtak}, R., {Gottl{\"o}ber}, S., \& {Klypin}, A. 2013, \mnras, 434, 1576

\bibitem[{{Wolf} \& {Bullock}(2012)}]{Wolf2012}
{Wolf}, J., \& {Bullock}, J.~S. 2012, ArXiv e-prints

\bibitem[{{Wolf} {et~al.}(2010){Wolf}, {Martinez}, {Bullock}, {Kaplinghat},
  {Geha}, {Mu{\~n}oz}, {Simon}, \& {Avedo}}]{Wolf2010}
{Wolf}, J., {Martinez}, G.~D., {Bullock}, J.~S., {Kaplinghat}, M., {Geha}, M.,
  {Mu{\~n}oz}, R.~R., {Simon}, J.~D., \& {Avedo}, F.~F. 2010, \mnras, 406, 1220

\bibitem[{{Zavala} {et~al.}(2013){Zavala}, {Vogelsberger}, \&
  {Walker}}]{Zavala2013}
{Zavala}, J., {Vogelsberger}, M., \& {Walker}, M.~G. 2013, \mnras, 431, L20

\bibitem[{{Zhao} {et~al.}(2003){Zhao}, {Jing}, {Mo}, \&
  {B{\"o}rner}}]{Zhao2003}
{Zhao}, D.~H., {Jing}, Y.~P., {Mo}, H.~J., \& {B{\"o}rner}, G. 2003, \apjl,
  597, L9

\bibitem[{{Zolotov} {et~al.}(2012){Zolotov}, {Brooks}, {Willman}, {Governato},
  {Pontzen}, {Christensen}, {Dekel}, {Quinn}, {Shen}, \&
  {Wadsley}}]{Zolotov2012}
{Zolotov}, A. {et~al.} 2012, \apj, 761, 71

\end{thebibliography}
\end{document}